\documentclass[aps,prd,twocolumn,superscriptaddress,nofootinbib,showpacs]{revtex4-1}
\usepackage[pass,letterpaper]{geometry}
\pdfoutput=1 
\usepackage[]{graphicx}
\usepackage{color}
\usepackage{latexsym,amsmath,amsfonts,amssymb,bm}
\usepackage{cancel}
\usepackage{booktabs}
\usepackage{verbatim}
\newcommand{\be}{\begin{equation}}
\newcommand{\ee}{\end{equation}}
\newcommand{\beq}{\begin{equation}}
\newcommand{\eeq}{\end{equation}}
\newcommand{\bea}{\begin{eqnarray}}
\newcommand{\eea}{\end{eqnarray}}

\newcommand{\dd}{\text{d}}

\newcommand\ees{\end{eqnarray}}
\newcommand\bees{\begin{eqnarray}}

\begin{document}

\title{Effects of a scalar fifth force on the dynamics of a charged particle as a new experimental design to test chameleon theories}

\author{Jean-Philippe Uzan}
\email{uzan@iap.fr}
\affiliation{Institut d'Astrophysique de Paris, CNRS UMR 7095,
Universit\'e Pierre \& Marie Curie - Paris VI, 98 bis Bd Arago, 75014 Paris, France}
\affiliation{Sorbonne Universit\'es, Institut Lagrange de Paris, 98 bis, Bd Arago, 75014 Paris, France}

\author{Martin Pernot-Borr\`as}
\email{martin.pernot\_borras@onera.fr}
\affiliation{DPHY, ONERA, Universit\'e Paris Saclay, F-92322 Ch\^atillon, France}
\affiliation{Institut d'Astrophysique de Paris, CNRS UMR 7095,
Universit\'e Pierre \& Marie Curie - Paris VI, 98 bis Bd Arago, 75014 Paris, France}

\author{Joel Berg\'e}
\email{joel.berge@onera.fr}
\affiliation{DPHY, ONERA, Universit\'e Paris Saclay, F-92322 Ch\^atillon, France}

\date{\today}
\begin{abstract}
This article describes the dynamics of a charge particle in a electromagnetic field in presence of a scalar fifth force. Focusing to the fifth force that would be induced by a chameleon field, the profile of which can be designed properly in the laboratory, it draws its physical effects on the cyclotron motion of a particle in a static and uniform magnetic field. The fifth force induces a drift of the trajectory that is estimated analytically and compared to numerical computations for profiles motivated by the ones of a chameleon field within two nested cylinders. The magnitude of the effect and the detectability of this drift are discussed to argue that this may offer a new experimental design to test small fifth force in the laboratory. More important, at the macroscopic level it induces a current that can in principle also be measured, and would even allow one to access the transverse profile of the scalar field within the cavity. In both cases, aligning the magnetic field with the local gravity field suppresses the effects of Newtonian gravity that would be several order larger than the ones of the fifth force otherwise and the Newtonian gravity of the cavity on the particle is also argued to be negligible. Given this insight, this experimental set-up, with its two effects -- on a single particle and at the macroscopic level -- may require attention to demonstrate its actual feasability in the laboratory.
\end{abstract}

\maketitle

\section{\bf Introduction.} 

The search for a fifth force of nature has a long history~\cite{Lee:1955vk,Fujii:1971vv,Gibbons:1981uz} related to the developments of the theories of gravitation beyond Newton and Einstein gravity. The existence of a scalar interaction~\cite{nord1912} has been revived by the development of theories of gravitation beyond general relativity since the existence of any new field may lead to a new long range force, depending on the nature of this new degree of freedom. 

Within the framework of scalar-tensor theories of gravitation~\cite{Damour:1993hw}, the extra scalar degree of freedom, $\phi$, is characterized by its potential $V(\phi)$ and its coupling to matter $A(\phi)$, so that the action of the theory, in the Einstein frame, is
\begin{equation}
\begin{split}
S = \int {\rm d}x^4 ~ &\sqrt{-g} \left[\frac{{\rm M}_{ \rm Pl}^2}{2}{\displaystyle R} -  \frac{1}{2}\partial^\mu\phi\partial_\mu\phi - V(\phi)\right]\\
&- \int {\rm d^4 }x \mathcal{L}_{\rm m}(\widetilde{g}_{\mu\nu}, {\rm matter})\sqrt{-\tilde g},
\end{split}
\end{equation}
with  ${\rm M}_{ \rm Pl}$ the reduced Planck mass, ${\displaystyle R}$ the Ricci scalar, $g_{\mu\nu}$ the Einstein frame metric, $g$ its determinant and $\mathcal{L}_{\rm m}$ the matter Lagrangian. The field couples non-minimally to matter through the Jordan frame metric $\widetilde{g}_{\mu\nu} = A^2(\phi)g_{\mu\nu}$, where $A(\phi)$ is a universal coupling function.

 If this field is massless, or its Compton wavelength is larger than the size of the Solar system, one can constrain its effects thanks to the parameterized post-Newtonian formalism~\cite{Will:2018bme,Will:2014kxa}. If the field is heavier, its action can be well-described by a Yukawa deviation from Newtonian gravity. Such Yukawa deviations, composition independent or dependent, have been tested from the sub-millimeter scales to the Solar system scales and cosmology~\cite{Adelberger:2003zx,Fischbach:1985tk,Fischbach:1999bc,Berge:2018htm,2013arXiv1309.5389J}, with recent stringent constraints obtained from the MICROSCOPE experiment~\cite{Berge:2017ovy,Touboul:2017grn}.

 If the coupling is universal then the scalar-tensor theory satisfies the weak equivalence principle. Besides, among those theories of gravity, General Relativity and Nordstr\"om theory which describes it by a scalar field in flat spacetime,  share the unique property to embody the strong equivalence principle; see e.g. Ref.~\cite{Deruelle:2011wu}. If the coupling is not universal, then the weak equivalence principle is violated and one expects a space-time variation of the fundamental constants, that can be tested in their own way~\cite{Uzan:2002vq,Uzan:2010pm,Uzan:2004qr}. Light scalar models can survive only if their coupling is extremely weak today, which can be ensured in a large class of models by an attraction mechanism toward general relativity during the cosmic history~\cite{Damour:1992kf,DAMOUR1994532}. Another class of models, including the symmetron~\cite{PhysRevLett.104.231301} and the chameleon \cite{khoury_chameleon_2004a,khoury_chameleon_2004} mechanisms, enjoy a screening mechanism in which the coupling or the mass of the field depends of the local matter density of matter. It follows that the environment can suppress the scalar force.

Many experimental set-ups have been proposed to test the chameleon mechanism in the laboratory, see Ref.~\cite{Weltman:2008fp,Steffen:2010ze,burrage_tests_2018,BraxReview,Berge:2017ovy} for reviews. This includes atomic spectroscopy~\cite{Brax:2010gp}; atom interferometry~\cite{burrageAtom, sabulsky_experiment_2018,burrage_probing_2015,elder_chameleon_2016,hamilton_atom-interferometry_2015,schlogel_probing_2016}, Casimir force measurement between plates ~\cite{brax_detecting_2007,burrage_proposed_2016}, that have extensively been used to test the inverse square law on sub-millimeter scales~\cite{Lamoreaux:2005zza,Lambrecht:2011qm};  the spectrum of ultra-cold neutrons in the Earth gravitational field~\cite{Brax:2011hb,PhysRevD.87.105013};  torsion balance experiments~\cite{Mota:2006ed,Adelberger:2006dh,upadhye_dark_2012}; neutron interferometry~\cite{Brax:2013cfa}. 

\vskip0.25cm
{\bf Goal.} The driving idea we want to investigate in this article, is to use the environmental set-up to design the profile of the scalar field inside the experiment, and hence the one of the fifth force. To that end, we rely on the results we obtained recently~\cite{PhysRevD.100.084006,us2} to determine the propagation of the chameleon field inside the MICROSCOPE satellite experiment. Hence, we have been able to compute the chameleon profiles (1) for one-dimensional systems made of parallel plates and (2) two-dimensional systems as inside a set of nested cylinders. Indeed when the axes of the cylinders are parallel but not coincident, hence shifted by $\delta$, the field distribution is no more cylindrically symmetric. It follows that the fifth force will modify the trajectory of any particle trapped between the cylinders. It is important to stress that in the chameleon situation, we can screen the experiment from the outside and design the profile of the fifth force inside the cavity. It will depend on the geometry of the cavity, the density inside the cavity and the parameters of the theory. This is a major difference with a light dilaton. The idea is thus to consider a charged particle in an electromagnetic field and determine the effect of the fifth force. Then, the system we shall consider is the trajectory of a particle orbiting inside two cylinders, or two parallel walls. This can be easily achieved thanks to a magnetic field. This latter case may offer an interesting set-up to design an experiment. The appendix give equations for the acceleration of a particle by an electric field in a capacitor with parallel walls, and adding a magnetic field, in order to determine if the fifth force affect the Hall tension. As we shall see, this does not offer an interesting method.

To that goal, we first derive in Section~\ref{section2} the general expression of the fifth force acting on a relativistic particle and its equation of motion in presence of an electromagnetic field; note that some subtleties concerning the fifth force have to be considered. We shall then focus in Section~\ref{section3} on the case of a static and uniform magnetic field and study the effect of the fifth force on the trajectory of the particle. As we shall explain, the fifth force induces a drift of the cyclotron motion with an amplitude and direction that depends on the characteristics of the fifth force. In Section~\ref{section4} we describe the macroscopic consequences of this drift. We will give estimates in order to discuss whether this can be measured and we will also compare it, in Section~\ref{section5}, to the reaction force arising from the radiation emitted by any charge particle. This will provide all the elements for discussion on the possibility to use such a set-up as a new experiment to constrain the existence of a fifth force. This analysis provides the first elements to discuss this possibility but also to estimate the possible effects of this scalar field on the propagation of high energy charged particle in the universe.

\vskip0.25cm
{\bf Set-up.} While most of our results will not depend on the specific choice of the coupling and the potential, let us be more specific on the choices that will be used for our numerics. We consider that the coupling function and potential  are of the form
\be\label{e.defA}
  A = \hbox{e}^{\beta \phi/{\rm M}_{\rm Pl}},
  \qquad
V= \Lambda^4\left(1+\frac{\Lambda^n}{\phi^n}\right)
\end{equation}
where $\Lambda$ is a mass scale, $n$ a natural number and $\beta$ a positive constant. It follows that the Klein-Gordon equation  involves an effective potential that depends on the  local the mass density $\rho$,
\begin{equation}\label{e.kg7} 
\Box\phi = \frac{{\rm d}V_{\rm eff}}{{\rm d}\phi},
\qquad
 V_{\rm eff}= V(\phi) + \frac{\beta}{{\rm M}_{\rm Pl}} \rho \, \phi.
\end{equation}
In our previous works, we have determined the profile of the scalar field for two parallel walls and two nested coaxial cylinders~\cite{PhysRevD.100.084006,us2} and when their axes is shifted~\cite{us2}. In the latter case, the profile is no more cylindrically symmetric so that a force appears between the two cylinders. In this work, we consider the trajectory of a particle of charge $q$ and mass $m$. 

Let us emphasize that the electromagnetic field does not modify the scalar field profile since external matter enters the Klein-Gordon equation only by a coupling to the trace of the stress-energy tensor through $T\ln A(\phi)$ in the effective potential~(\ref{e.kg7}). We consider the Cartesian basis $({\bm e}_x,{\bm e}_y,{\bm e}_z)$ and cylindrical basis $({\bm e}_r,{\bm e}_\theta,{\bm e}_z)$ aligned with the magnetic field.\\

\section{Dynamics of a charged particle}\label{section2}

\subsection{\bf Fifth force.}  

Since the matter fields couple to the metric $A^2(\phi)g_{\mu\nu}$ the equation of a point particle of mass $m$ and charge $q$ derives from the action
\be\label{e.action}
S_{\rm pp}=-c^2\int m(\phi)\sqrt{-g_{\mu\nu}u^\mu u^\nu}\dd\tau+ q \int A_\mu u^\mu \dd\tau
\ee
where $\tau$ is the proper time and $u^\mu$ the tangent vector to  the worldline, i.e.  $u^\mu=\dd X^\mu/\dd\tau$ and satisfies $u_\mu u^\mu =-c^2$ and $A_\mu$ the potential vector. Since we are considering particle, i.e. weakly self-gravitating bodies, the mass functions $m(\phi)$ reduces to $mA(\phi)$ with $m$ constant, the Jordan mass, such that particles with $q=0$ follow geodesic of the metric $\tilde g_{\mu\nu}$. The equation of motion is
\begin{eqnarray}\label{e.motion}
mc^2 \gamma^\mu = \frac{q}{A(\phi)}{F^\mu}_\nu u^\nu - mc^2\frac{\partial \ln A}{\partial\phi}\perp^{\mu\nu}\nabla_\nu\phi
\end{eqnarray}
with $\gamma^\mu\equiv u^\nu\nabla_\nu u^\mu=\dd u^\mu/\dd\tau$ is the 4-acceleration and satisfies $\gamma^\mu u_\mu=0$,  $F_{\mu\nu} = \partial_\mu  A_\nu -\partial_\nu A_\mu$ the Faraday tensor, $\perp_{\mu\nu}\equiv g_{\mu\nu}+u_\mu u_\nu/c^2$ the projector on the 3-space normal to $u^\mu$, which indeed ensures that $u^\mu u_\mu=-c^2$; see e.g. Ref.~\cite{Uzan:2010pm}. It follows that the fifth force,
\be
F^\mu =  - mc^2\frac{\beta(\phi)}{M_{\rm P}}\perp^{\mu\nu}\nabla_\nu\phi,
\ee 
remains perpendicular to the 4-velocity, $u_\mu F^\mu=0$. Indeed, this equation is 4-dimensional and we shall see below that in the 3-dimensional langage, it is associated to a non-vanishing work. $\beta$, defined by
\be\label{e.defbeta}
\beta(\phi)=M_{\rm P}\frac{\dd \ln A}{\dd\phi},
\ee
characterizes the sensitivity of the mass to a variation of the scalar field; it is dimensionless. Clearly, in the Galilean limit the projector plays no role.  Note also that the Lorentz force is proportional to $q/A(\phi)$, the factor $A$ arising from the fact that the Einstein mass is $mA(\phi)$. From now on, we work in units in which $c=1$.

\subsection{\bf Equations of motion.}  In the Newtonian limit $g_{\mu\nu}$ reduces to the Minkowski metric $\eta_{\mu\nu}$ and the geodesic is given in 3-dimensional notations $X^\mu=(T,{\bm X})$. We define the 3-velocity and 3-acceleration as
\be
{\bm V}=\frac{\dd{\bm X}}{\dd T}, \quad
{\bm a}=\frac{\dd{\bm V}}{\dd T},
\ee
where we use the convention that ${\bm V}$ have coordinates $V^i$ with $i=1\ldots3$. With these notations (see Ref.~\cite{DUbook} for details), the scalar product is indeed ${\bm a}.{\bm V}=\delta_{ij} a^iV^j$ and we have
$$
u^0=\frac{\dd X^0}{\dd \tau} = \frac{1}{\sqrt{1-V^2}},\quad
u^i=\frac{\dd X^i}{\dd \tau} = \frac{V^i}{\sqrt{1-V^2}}
$$
with $V^2=\delta_{ij}V^iV^j$ and
$$
\gamma^0= \frac{{\bm a}.{\bm V}}{(1-V^2)^2},\quad
\gamma^i= \frac{1}{1-V^2}\left( a^i + \frac{{\bm a}.{\bm V}}{1-V^2} V^i \right).
$$
The scalar force reduces to the Nordstr\"om force~\cite{nord1912} (see also \S~10.3 of Ref.~\cite{DUbook}) and, once the Faraday tensor is decomposed as $F_{0i}=-E_i$, $F_{jk}=e_{ijk}B^i$ with $e_{ijk}$ the Levi-Civita symbol, the Lorentz force has components
\be
F^0_L = \frac{q/A(\phi)}{\sqrt{1-V^2}} {\bm E}.{\bm V}, \quad
{\bm F}_L= \frac{q/A(\phi)}{\sqrt{1-V^2}}\left( {\bm E}+ {\bm V}\wedge{\bm B}\right)
\ee
and the equation of motion~(\ref{e.motion}) splits as
\begin{widetext}
\begin{eqnarray}
&& \frac{{m\bf a}.{\bm V}}{(1-V^2)^2} =\frac{q/A(\phi)}{\sqrt{1-V^2}} {\bm E}.{\bm V}- \frac{m}{M_{\rm P}}\beta \frac{{\bm V}.\nabla\phi}{1-V^2},\label{e.a0}\\
&&\frac{m}{1-V^2}\left( {\bm a}+ \frac{{\bm a}.{\bm V}}{1-V^2} {\bm V}  \right)= \frac{q/A(\phi)}{\sqrt{1-V^2}}\left[ {\bm E}+ {\bm V}\wedge{\bm B} \right] - \frac{m}{M_{\rm P}} \beta\left(\nabla\phi +\frac{({\bm V}.\nabla\phi)}{1-V^2} {\bm V}\right),\label{e.ai}
\end{eqnarray}
respectively for the time  and space components. Eq.~(\ref{e.ai}) can be rewritten in a more compact form as
\begin{eqnarray}
&&\frac{\dd}{\dd T}\left(\frac{m{\bm V}}{\sqrt{1-V^2}} \right) = \frac{q}{A(\phi)}\left[ {\bm E}+ {\bm V}\wedge{\bm B} \right] - \frac{m}{M_{\rm P}} \beta\sqrt{1-V^2}\left(\nabla\phi +\frac{({\bm V}.\nabla\phi)}{1-V^2} {\bm V}\right).\label{e.ai2}
\end{eqnarray}
This form makes explicit the 3-momentum ${\bm P}\equiv m{\bm V}/\sqrt{1-V^2}$ so that the r.h.s. is just the sum of the 3-dimensional form of the 2 forces, ${\bm f}_{\rm em}+{\bm f}_5$. Note also that once multiplied by ${\bm V}$ and using Eq.~(\ref{e.a0}), it takes a form closer to the standard Newton third law,
\begin{eqnarray}
&&\frac{m{\bm a}}{1-V^2}= \frac{q/A(\phi)}{\sqrt{1-V^2}}\left[ {\bm E}+ {\bm V}\wedge{\bm B} -({\bm E}.{\bm V}){\bm V}\right] - \frac{m}{M_{\rm P}} \beta\nabla\phi.\label{e.adyn}
\end{eqnarray}
\end{widetext}
This provides the general relativistic equations of propagation of a charged particle in an electromagnetic field in presence of a fifth force. 

\subsection{\bf Conservation of energy} 

For a static field with $A^\mu=(\Phi_E,{\bm A})$, it is easily checked that Eq.~(\ref{e.a0}), with  use of the definition~(\ref{e.defbeta}), implies that
\be
\frac{\dd}{\dd T}\left[\frac{m A(\phi)}{\sqrt{1-V^2}}+q\Phi_E\right]=0
\ee
for any static configuration of the fields,  hence the conservation of the energy of the particle
\be\label{e.defE}
{\cal E}\equiv \frac{m A(\phi)}{\sqrt{1-V^2}}+q\Phi_E.
\ee

The point particle action is easily rewritten as $\int {\cal L}\dd T$ defining the Lagrangian
\be
{\cal L}= -m A(\phi)\sqrt{1-V^2}+ q{\bm A}.{\bm V} -q\Phi_E
\ee
from which we deduce the conjugate momenta
\be\label{e.defpi}
{\bm\pi}=\frac{m A(\phi)}{\sqrt{1-V^2}} {\bm V} + q{\bm A}.
\ee
Indeed, the Hamiltonian ${\cal H}={\bm\pi}.{\bm V}-{\cal L}$ reduces to the expression~(\ref{e.defE}) of the energy or equivalently to
\be
{\cal H}=\sqrt{m^2A^2(\phi) + ({\bf \pi}-q{\bm A})^2} + q\Phi_E.
\ee
As we shall see, the Lagrange equations 
$$
\frac{\dd {\bm\pi}}{\dd T}= {\bm\nabla}{\cal L}
$$
will provide additional conserved quantities once the symmetries of the problem are specified.

\section{Particle in a magnetic field}\label{section3}

We now assume that the particle is subject to a static and uniform magnetic field,  parallel to the axis of the cylinders, ${\bm B}=B{\bm e}_z$. It follows that
\be
{\bm A}({\bf r}) = \frac{1}{2}{\bm B}\wedge{\bf r} =\frac{1}{2}Br{\bm e}_\theta
\ee
and the cyclotron pulsation
\be\label{e.defw0}
 \omega_0 =\frac{qB}{m},
 \ee
is of the order of
\be\label{e.w0}
 \omega_0 = 9.5\times10^7\, Z \left(\frac{B}{1\,{\rm T}}\right)\left(\frac{m}{m_p}\right)^{-1}\,{\rm s}^{-1},
\ee
$m_p$ being the proton mass and $Z$ the charge number.

\subsection{Cyclotron motion}

When the fifth force vanishes, the equations of motion are easily integrated to give
\be
\frac{\dd u^{0,3}}{\dd\tau}= 0, \quad
\frac{\dd u^1}{\dd\tau}= w_0 u^2, \quad
\frac{\dd u^2}{\dd\tau}= -w_0 u^1, 
\ee
the solution of which  is
\begin{eqnarray}\label{e.7}
&&X=R_0\sin \omega_0\tau, \qquad
Y=R_0\cos \omega_0\tau, \nonumber \\
&&Z= U_Z\tau,\qquad\qquad
T-T_0 = \frac{\omega_0}{\Omega}\tau.
\end{eqnarray}
with $U_Z$, $T_0$ and $R_0$ constants of integration and
\be\label{e.18}
\Omega = \frac{\omega_0}{\sqrt{1+R_0^2\omega_0^2+U_Z^2}}.
\ee
The charge travels on an helix of radius $R_0$ and pitch $2\pi U_Z/\omega_0$ about ${\bm B}$ with an angular velocity ${\omega_0}$ (the cyclotron frequency) when measured with its proper time and ${\Omega}$ (the synchrotron frequency) when measured with the coordinate time $T$ of the inertial frame. Note that since  $V^2=(R_0^2\omega_0^2+U_Z^2)/(1+R_0^2\omega_0^2+U_Z^2)$ we have $\Omega = \omega_0\sqrt{1-V^2}$. We deduce that the Larmor radius is given by
\be
R_0=\frac{V}{\omega_0\sqrt{1-V^2}}\sin\psi, \quad
U_Z = \frac{V}{\sqrt{1-V^2}}\cos\psi,
\ee
$\psi$ being the pitch angle.

\subsection{Conserved quantities}

In the configuration considered here, the electric field vanishes and the magnetic field has been chosen as ${\bm B}=B{\bm e}_z$ and the field configuration as $\phi(x,y)$. It follows from Eq.~(\ref{e.adyn}) that 
\be
{\bm a}.{\bm e}_z=0
\ee
 so that $V_z$ remains constant. In the following we shall assume $V_z=0$ so that the motion is reduced to a plane perpendicular to $z$. 
 
 Then, Eq.~(\ref{e.defpi}) implies that the motion satisfies the constraint
\be
\frac{\dd\pi_\theta}{\dd T}=-m A(\phi) \sqrt{1-V^2}\partial_\theta\ln A
\ee
so that $\pi_\theta$, given by
\be\label{e.defL}
\pi_\theta= mr^2\left(\frac{A(\phi)\dot\theta}{\sqrt{1-V^2}} + \frac{1}{2}\omega_0\right),
\ee
can be identified with the angular momentum and conserved if $\phi$ has an axial symmetry, i.e. if the fifth force is radial. We also recall that the energy
\be\label{e.defE}
{\cal E}=\frac{m A(\phi)}{\sqrt{1-V^2}}
\ee
will be conserved.

\subsection{Non-relativistic cyclotron motion.}  

\subsubsection{Non-relativist equations}

In the non-relativistic regime the equation of motion~(\ref{e.adyn}) reduces to
$$
{\bm a} = \frac{\omega_0}{A(\phi)} {\bm v}\wedge{\bm e}_z - \beta\nabla\varphi,
$$
with $\varphi=\phi/M_{\rm P}$. Even if the gradient of $\phi$ can be important, $A$ remains close to unity because $\varphi \ll 1$ (see Fig.~\ref{Fig8} below for a concrete numerical example). So we shall approximate the dynamics by
\be\label{e.1}
{\bm a} = \omega_0 {\bm v}\wedge{\bm e}_z - \beta\nabla\varphi,
\quad\hbox{with}\quad
 \omega_0 = \frac{qB}{m},
\ee
i.e. $A\sim A_0=1$. We assume that the two cylinders have axis parallel to ${\bm e}_z$ so that the scalar field profile is independent of $z$, i.e. $\phi(x,y)$ or $\phi(r,\theta)$ in either Cartesian coordinates or cylindrical coordinates. Hence, we got the system
\be\label{e.1b}
\left\lbrace
\begin{array}{ccc}
\ddot x  &=& \omega_0 \dot y  - \beta c^2\partial_x\varphi \\
 \ddot y &=& -\omega_0 \dot x  -\beta c^2 \partial_y\varphi
\end{array}
\right. .
\ee
It can trivially be checked that the conserved quantities reduce to
\be\label{e.E2}
{\cal E}=\frac{1}{2}(\dot x^2+\dot y^2)+\beta c^2\varphi
            = \frac{1}{2}(\dot r^2+r^2\dot \theta^2)+\beta c^2\varphi
\ee
for the massic energy~(\ref{e.defE}), that is indeed conserved and the massic angular momentum~(\ref{e.defL})
\be\label{e.L2}
\ell_z = r^2\left(\dot\theta+\frac{1}{2}\omega_0\right)
\ee
is conserved only for cylindrically symmetric field configuration since
\be\label{e.L2b}
\dot \ell_z = \beta c^2\partial_\theta\varphi.
\ee

\subsubsection{\it Orders of magnitude.} 

To put some numbers, the pulsation is given by Eq.~(\ref{e.w0}) so that the radius of the trajectory in absence of a fifth force is
\be
R_0 = 1.4\times10^{-4}  \left(\frac{E_0}{1\,{\rm eV}}\right)^{1/2}  \left(\frac{m}{m_p}\right)^{1/2}Z^{-1} \left(\frac{B}{1\,{\rm T}}\right)^{-1}
\,{\rm m}.
\ee

\subsubsection{\it Dynamics with no fifth force}  We have already discussed the free motion in full generality. We just need to add the connection to the initial conditions and consider a new description of the motion. 

Assume that at $t=0$ the trajectory starts at $(x_0,y_0)$ with velocity $(V_0\cos\alpha, V_0\sin\alpha)$, its equation is then
\be\label{e.22}
\left\lbrace
\begin{array}{ccc}
x  &=&x_c + R_0\sin\left(\omega_0 t-\alpha\right)  \\
y &=&y_c + R_0\cos\left(\omega_0 t-\alpha\right)
\end{array}
\right.
\ee
with
\be
R_0= V_0/\omega_0,\qquad
\left\lbrace
\begin{array}{ccc}
x_c  &=&x_0 + R_0\sin\alpha  \\
y_c &=&y_0 - R_0\cos\alpha
\end{array}
\right..
\ee
$R_0$ can be negative with our convention. This is indeed trivial but it emphasizes that the center of the motion is not the center of the coordinates system because the magnetic force is not a central force. It is easily checked that
$$
\ell_z=\frac{1}{2}\left( r_c^2-R_0^2\right)\omega_0,\qquad
{\cal E}=\frac{1}{2}R_0^2\omega_0^2
$$
so that ${\cal E}=V_0^2/2$ gives the relation between the radius of the orbit and the pulsation. 

Since
\be\label{e.cons00}
\left\lbrace
\begin{array}{ccl}
\frac{\dd\theta}{\dd t}&=&\frac{\ell_z}{r^2} -\frac{1}{2}\omega_0\\
\left(\frac{\dd r}{\dd t}\right)^2 &=& 2{\cal E} - r^2\left(\frac{\ell_z}{r^2}-\frac{1}{2}\omega_0 \right)^2
\end{array}
\right.,
\ee
the minimum and maximum radius of the trajectory are given by
$$
r_\pm = \frac{\sqrt{2}}{\omega_0}\sqrt{2{\cal E} + \ell_z\omega_0\pm2\sqrt{{\cal E}({\cal E}+\ell_z\omega_0)}}
$$
that satisfy $r_+-r_-=2R_0$ as expected. Now, obviously $\theta$ is not constant so that the period of the motion cannot be extracted directly, however, since $\dd t= \dd r /\dot r$, we have from Eq.~(\ref{e.cons00}) that
$$
 t =\int\frac{\dd r}{\sqrt{2{\cal E} - r^2\left(\frac{\ell_z}{r^2}-\frac{1}{2}\omega_0 \right)^2}},
$$
so that the period of the motion is
\be\label{e.period}
 \frac{T}{2} =\int_{r_-}^{r_+} \frac{2 r\dd r/\omega_0}{\sqrt{(r_-^2-r^2)(r^2-r_+^2)}}=\frac{\pi}{\omega_0}.
\ee
Note also that Eq.~(\ref{e.cons00}) shows that the dynamics is the one of a point particle with a potentiel $(\omega_0^2r^2/4-\ell_z\omega_0)/2$, that is nothing but the centrifugal potential. This may sound as a complicated way of describing a simple result but this can be easily generalized to the case of a perturbing force.

\subsubsection{\it Radial force} 

\begin{figure}[htb]
\centering
 \includegraphics[scale= 0.4]{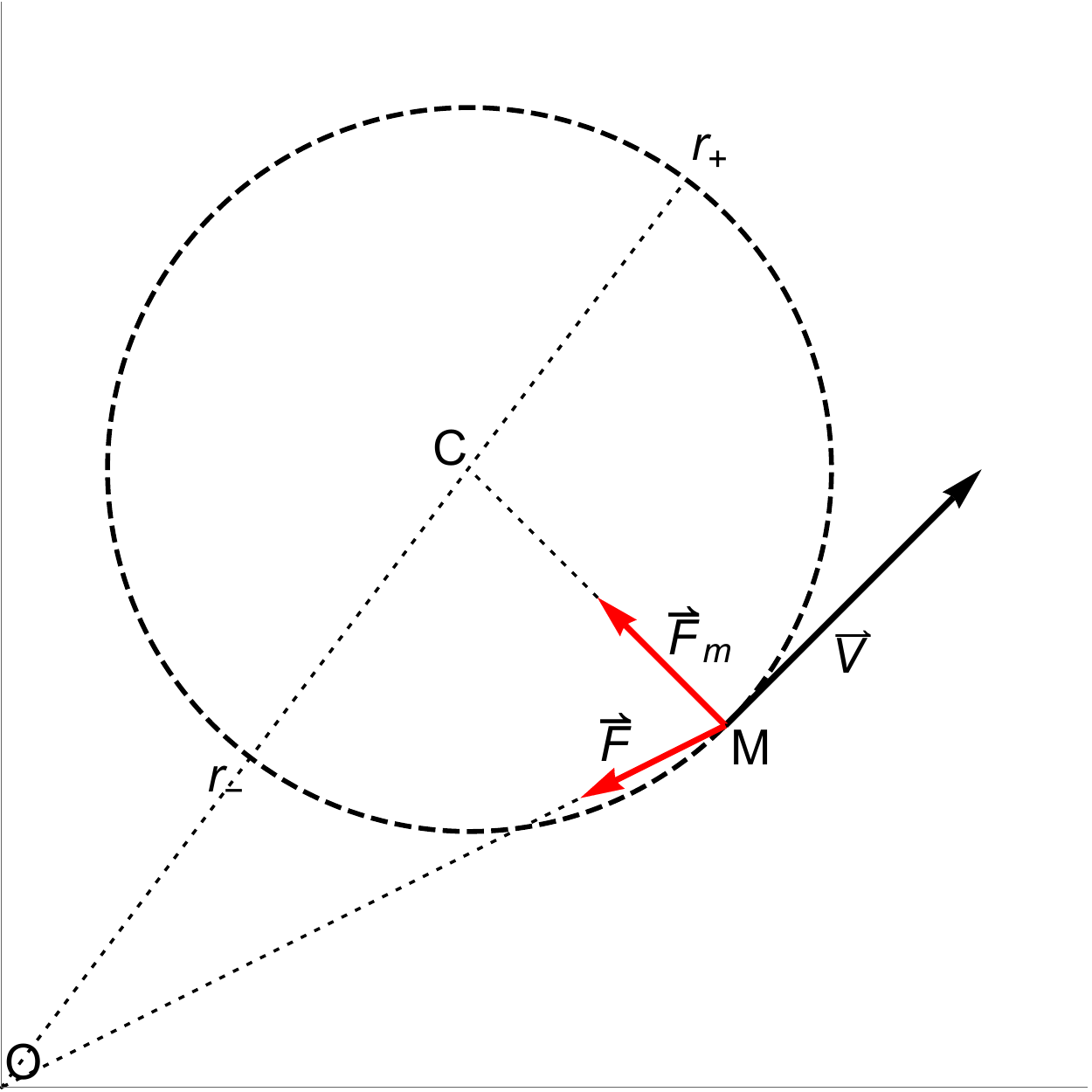}
   \caption{The geometry of the problem. The magnetic force ${\bm F}_m$ is perpendicular to the motion and thus points locally toward the center of curvature $C$ of the trajectory. The dashed circle represents the pure magnetic trajectory. The perturbative force ${\bm F}$, even is central is not parallel to ${\bm F}_m$ unless $C=O$. We call ``radial" a force for which there exists a coordinates system such that $\phi(r)$, i.e. such that the force points toward $O$. the magnetic force points toward the local curvature center and is thus not radial but simply perpendicular to the trajectory.}
   \label{Fig1} 
\end{figure}

As can be trivially seen from Fig.~\ref{Fig1}, the magnetic force is indeed not radial. It points toward the local center of curvature. With our definition $\phi/M_{\rm P}=\varphi(r)$ the force per unit mass is  ${\bm F}=\beta c^2 \varphi'(r){\bm e}_r$. Since the field enjoys a cylindrical symmetry, the angular momentum~(\ref{e.L2}-\ref{e.L2b}) is conserved. We deduce that
\be\label{e.cons01}
\left\lbrace
\begin{array}{ccl}
\frac{\dd\theta}{\dd t}&=&\frac{\ell_z}{r^2} -\frac{1}{2}\omega_0\\
\left(\frac{\dd r}{\dd t}\right)^2 &=& 2{\cal E}- r^2\left(\frac{\ell_z}{r^2}-\frac{1}{2}\omega_0 \right)^2 -2\beta c^2\varphi(r) 
\end{array}
\right.,
\ee
which is a simple extension of Eq.~(\ref{e.cons00}). This shows that the dynamics is similar to the one of a point particle of unit mass in the effective potential
$$
U_{\rm eff}= \beta \varphi(r)+\frac{r^2}{2}\left(\frac{\ell_z}{r^2}-\frac{1}{2}\omega_0\right)^2.
$$
The integration of Eq.~(\ref{e.cons01}) by quadrature gives
\begin{eqnarray}
 t &=&\int\frac{\dd r}{\sqrt{2({\cal E} -\beta c^2\varphi(r)  )- r^2\left(\frac{\ell_z}{r^2}-\frac{1}{2}\omega_0 \right)^2 }},\label{e.gen}\\
 \theta &=&\int\frac{(\ell_z/r^2-\omega_0/2)\dd r}{\sqrt{2({\cal E} -\beta c^2\varphi(r)  )- r^2\left(\frac{\ell_z}{r^2}-\frac{1}{2}\omega_0 \right)^2}}
\end{eqnarray}
which gives the equation of the trajectory in the parametric form $\lbrace t(r),\theta(r)\rbrace$.  The turning points are solution of
\be\label{e.annulus}
\dot r =0.
\ee
They delimit the domain of the allowed motion. If this domain is of the form $[r_-,r_+]$ then the trajectory is restricted to an annulus and, thanks to Bertrand theorem (1873), we know that the trajectory will be periodic only of $\varphi\propto r^2$ or $1/r$.

\begin{figure}[htb]
\centering
 \includegraphics[scale= 0.35]{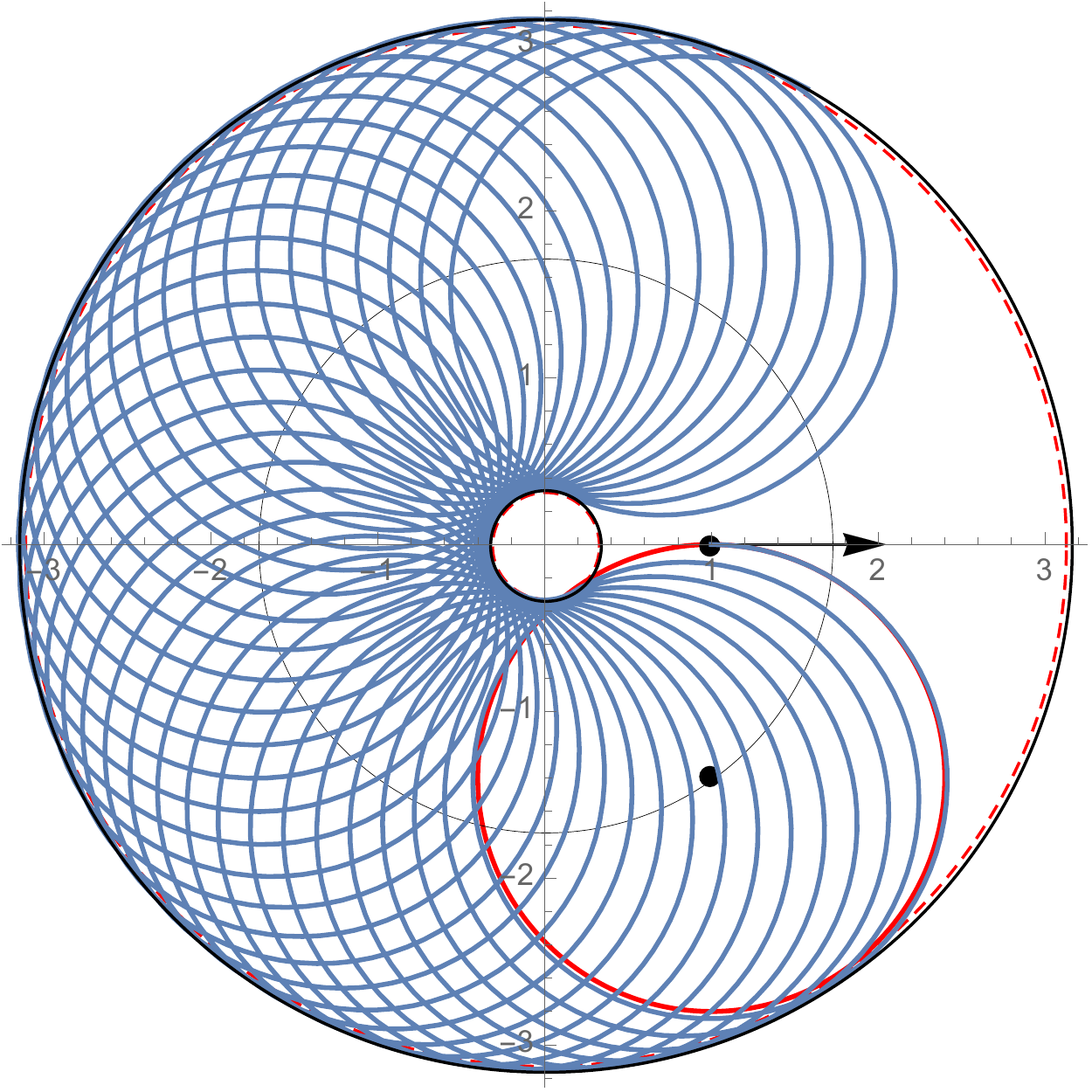} \includegraphics[scale= 0.35]{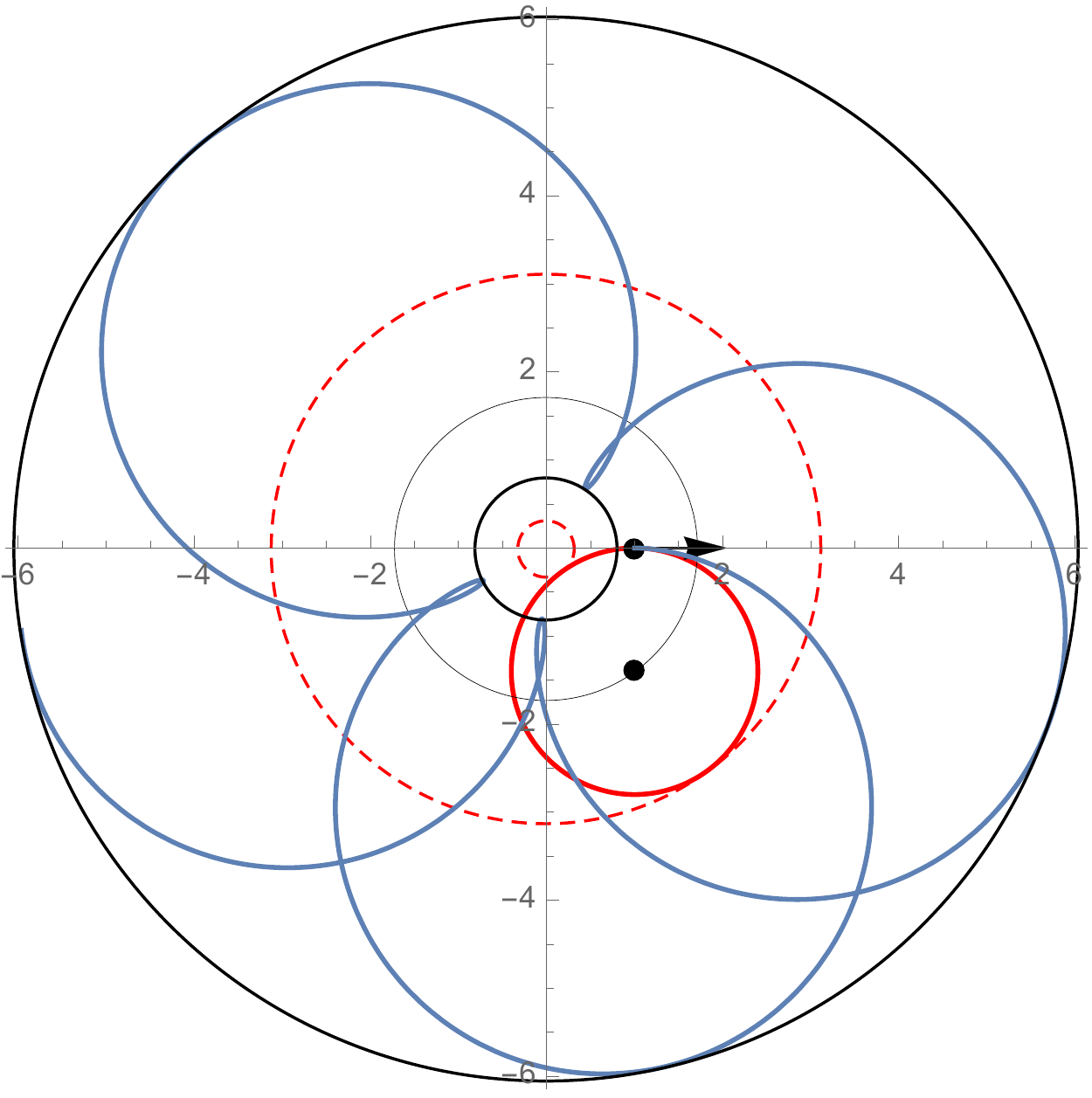} \includegraphics[scale= 0.35]{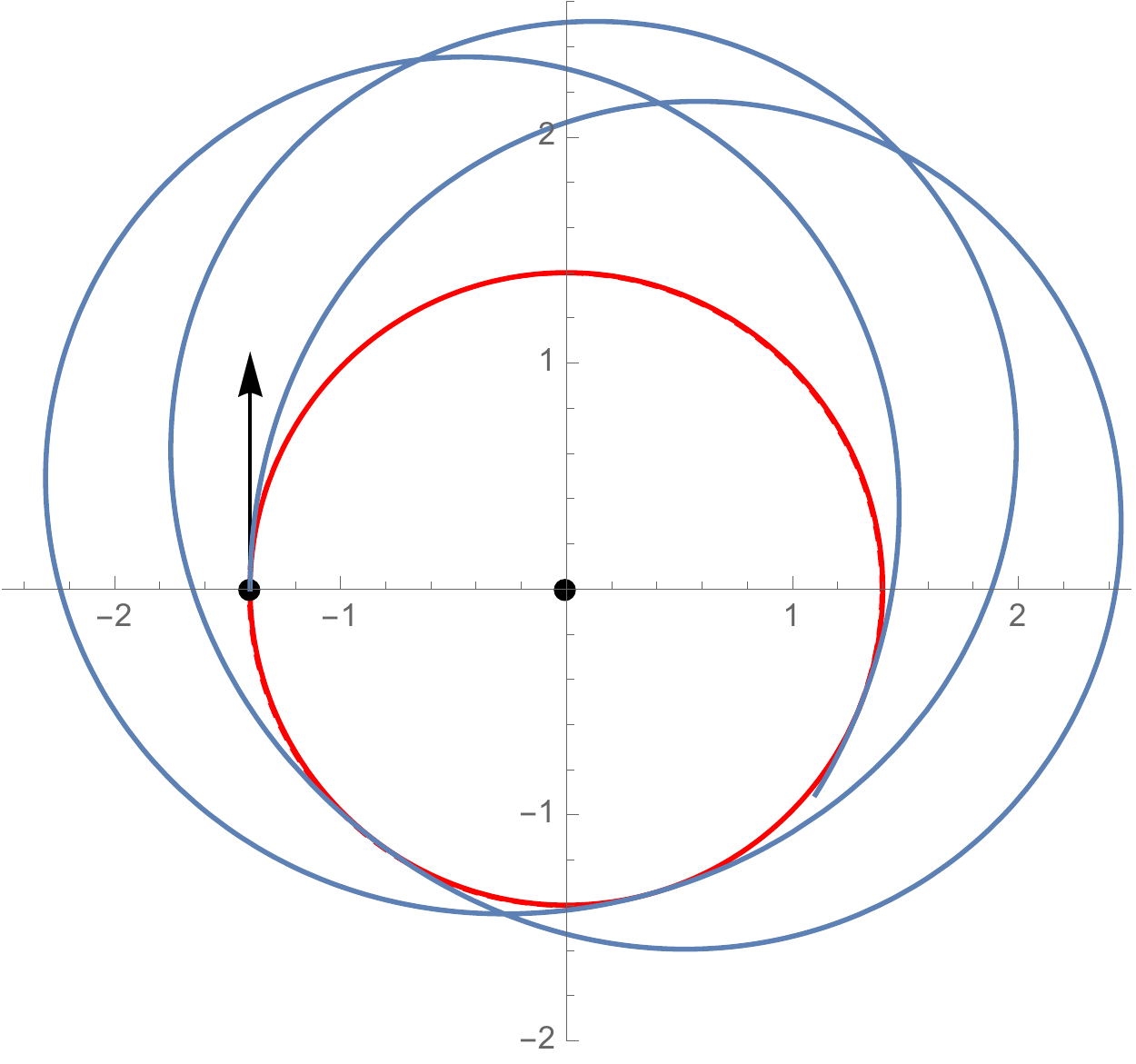}
   \caption{Example of a central force $\varphi=a/r$. The solid red circle corresponds to the free motion ($a=0$) while the dashed red circles define the annulus of allowed trajectory when there is no fifth force. We have represented the initial conditions (initial point and initial velocity) as well as the center of the magnetic trajectory (black dot). When $a\not=0$ the two black circles represent the turning radii defining the annulus of allowed trajectories. When $\beta a c^2$ is small (top: $\beta a c^2=0.1$~m$^3$/s$^{2}$) the free trajectory precesses slowly inside this annulus. When $a$ is larger (middle, $\beta a c^2=1$~m$^3$/s$^{2}$), the trajectory can explore regions forbidden in absence of the fifth force. The last example considers the case in which the center of the free magnetic motion is $O$ so that it will the static circular trajectory is deformed in a precessing ellipse ($\beta a c^2=-0.3$~m$^3$/s$^{2}$). All plots assume $\omega_0=0.5$~s$^{-1}$, $V_0=0.7$~m/s and $x_0=1$~m.}
   \label{Fig2} 
\end{figure}

Numerically, once we set the initial conditions $(x_0,y_0)$ and $V_0(\cos\alpha,\sin\alpha)$ it is obvious that $r_0=\sqrt{x_0^2+y_0^2}$, $\theta_0=\arctan(y_0/x_0)$, $\dot\theta_0=(V_0/r_0)\sin(\alpha-\theta_0)$ so that the energy and angular momentum are ${\cal E}=V_0^2/2+\beta c^2\varphi(r_0)$ and $\ell_z=r_0^2(\dot\theta_0+\omega_0/2)$, which determines the annulus of allowed trajectories. As an example, we consider the potential $\varphi=a/r$, with $a$ a constant with units of length. When $a\rightarrow 0$ we recover the free trajectory which is then drifting along the center defined by the central force, as shown on Fig.~\ref{Fig2} (the value of the parameters are not meant to be realistic but chosen to illustrate the properties of the trajectory). Note also that by tuning the initial conditions, we can either get a small  trajectory drifting in between the cylinders or a large trajectory precessing  around the inner cylinder.

\begin{figure}[htb]
\centering
 \includegraphics[scale= 0.3]{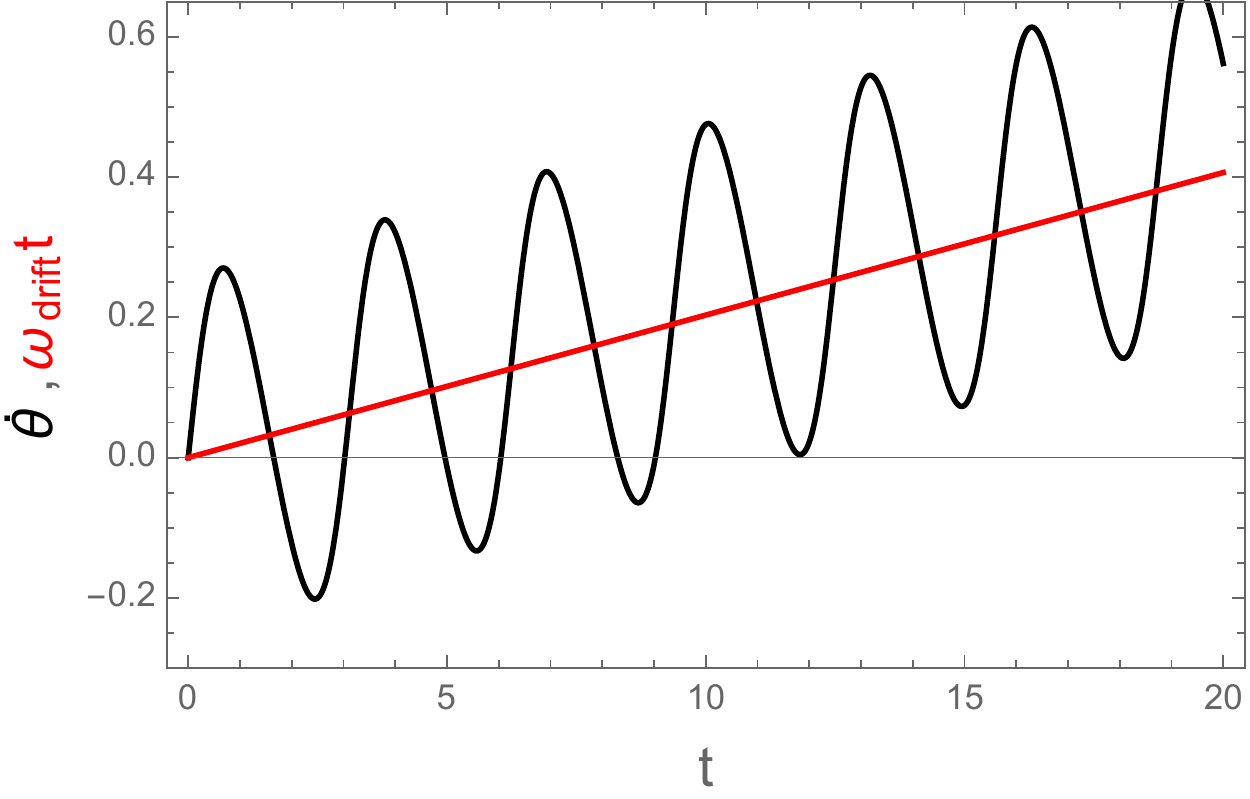} \includegraphics[scale= 0.3]{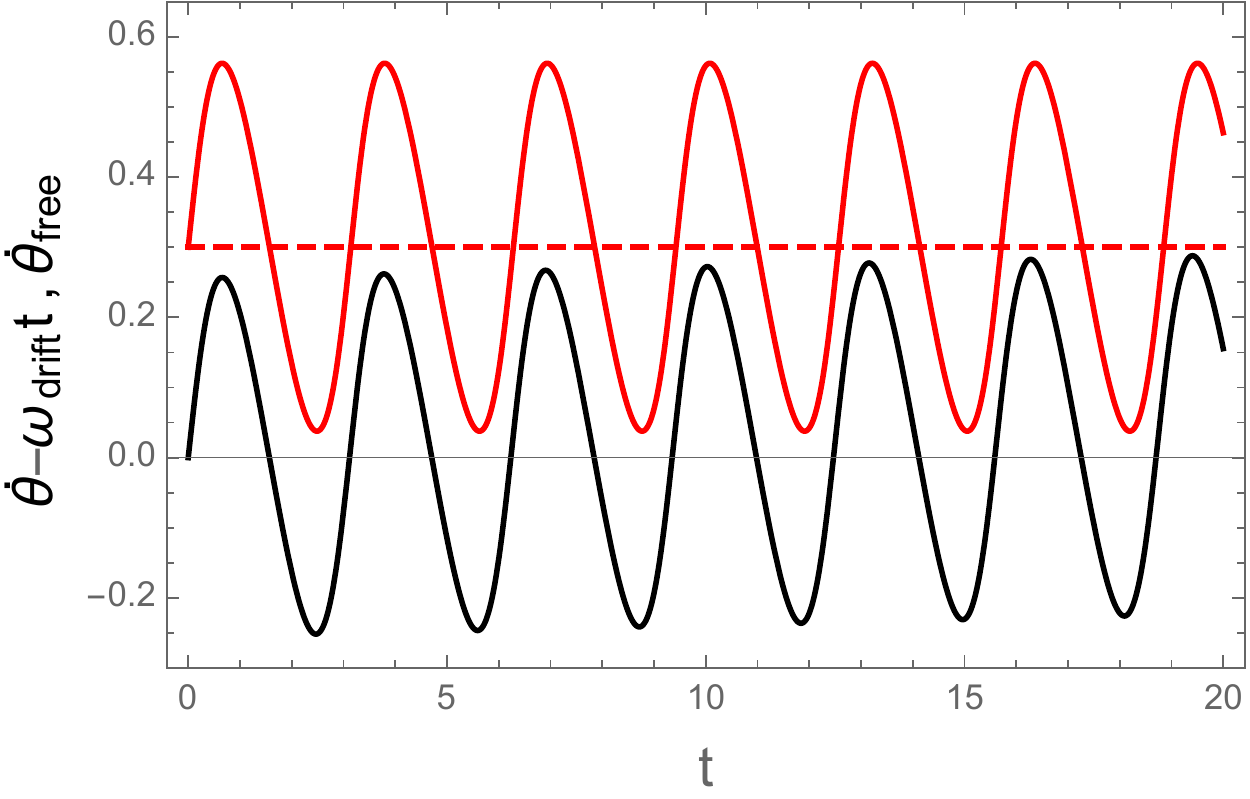}
   \caption{Comparison of the variation of $\theta(t)$ with the drift $\omega_{\rm drift}t$ [left] and of the residue $\theta-\omega_{\rm drift}t$ to the $\theta(t)$ for the free motion with an arbitrary offset to compare the curves [right]. Parameters: $\omega_0=2$, $a=0.1$, $V_0=0.7$, $x_0=1$~m.}
   \label{Fig3} 
\end{figure}

Since the fifth force is small compared to the magnetic force, we can estimate the period of the drift from the fact that in the guiding center approximation~\cite{driftart}, the drift velocity is 
\be\label{e.drift}
{\bm v}_{\rm drift}= \frac{{\bm F}\wedge{\bm B}}{q B^2},
\ee
which holds as long as the force can be consider constant on the scale of the gyroradius, a condition that is satisfied for our models. For a radial force $-m\beta c^2\varphi'{\bm e}_r$ and a magnetic field along ${\bm e}_z$ this leads to an orthoradial velocity,
\be\label{e.driftr}
{\bm v}_{\rm drift} = \frac{\beta c^2\varphi'}{\omega_0}{\bm e}_\theta
\ee
that is to the pulsation of the drift of the trajectory of $C$ around $O$ as
\be
\omega_{\rm drift} = \left.\frac{\beta c^2\varphi'}{r\omega_0}\right\vert_{r=r_c}.
\ee
This is indeed an approximation which works well when the force is small and when the gradient of the fifth force is small on the scale of the gyroradius. Fig.~\ref{Fig3} shows that it gives an excellent estimation of the drift pulsation.

\subsubsection{\it Radial chameleon}  

The previous analysis shows that a tiny central force will modify the cyclotron motion in two ways: (1) by extending the zone of allowed trajectories and (2) by making the trajectory drift. For a fifth force of small amplitude we are mostly interested by the latter effect. 

The advantage of the chameleon field is that we can ``engineer" the profile of the field inside the cavity.  If the two cylinders have the same axis, then the experiment enjoys a cylindrical symmetry and $\varphi(r)$ so that the fifth force is radial. The simulations we are using~\cite{PhysRevD.100.084006,us2} assume that $R_{\rm in}=0.2$~m, $R_{\rm out}=0.6$~m, $\rho_{\rm mat}=8.125~{\rm g.cm}^{-3}$ (typical of invar) for the cylinders and $\rho_{\rm in}=10^{-3}\rho$ for the inter-cylinder region. The theory assumes $\Lambda=1$~eV, $n=2$, $\beta=1$.

\begin{figure}[htb]
\centering
 \includegraphics[scale= 0.3]{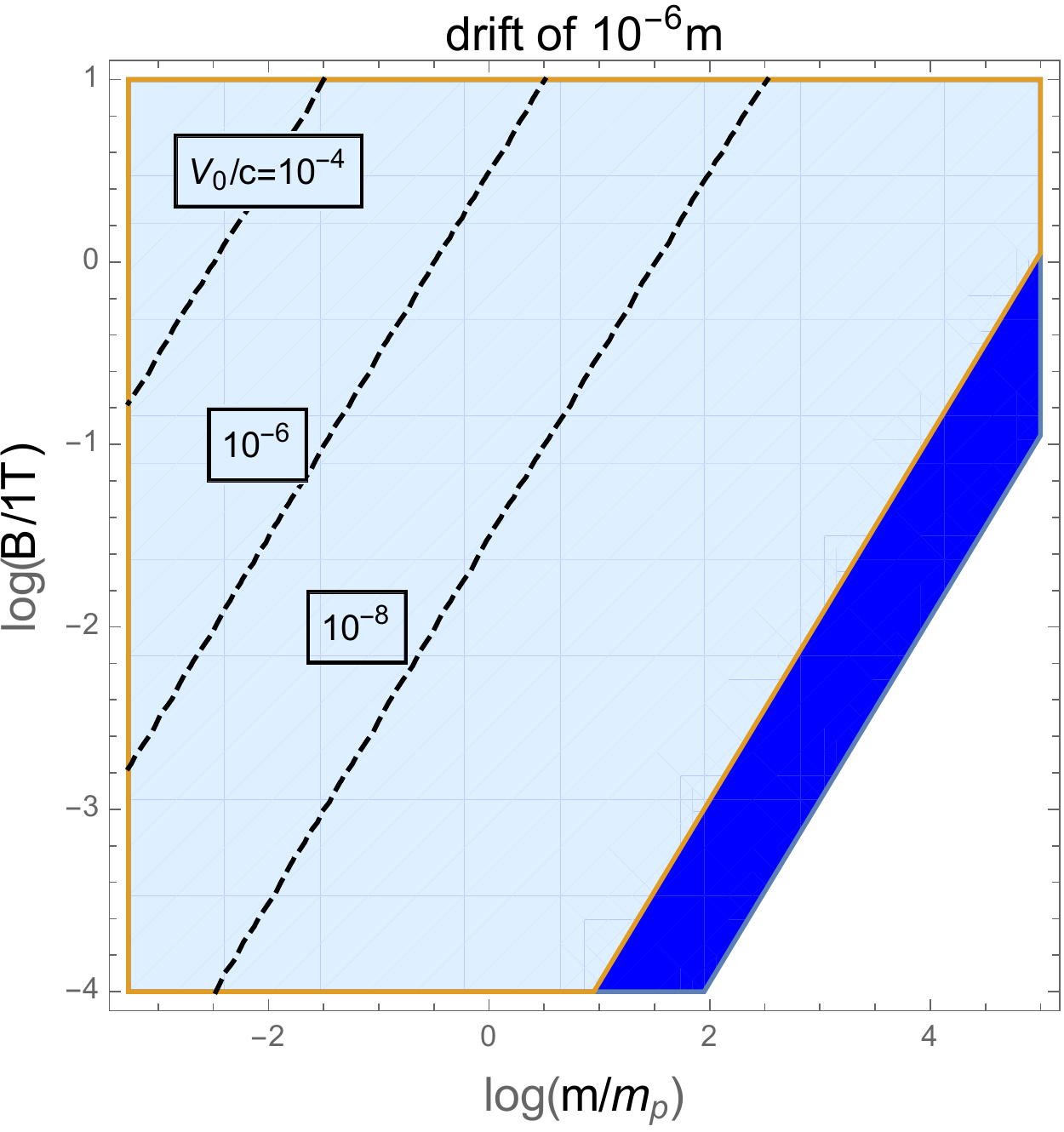} \includegraphics[scale= 0.3]{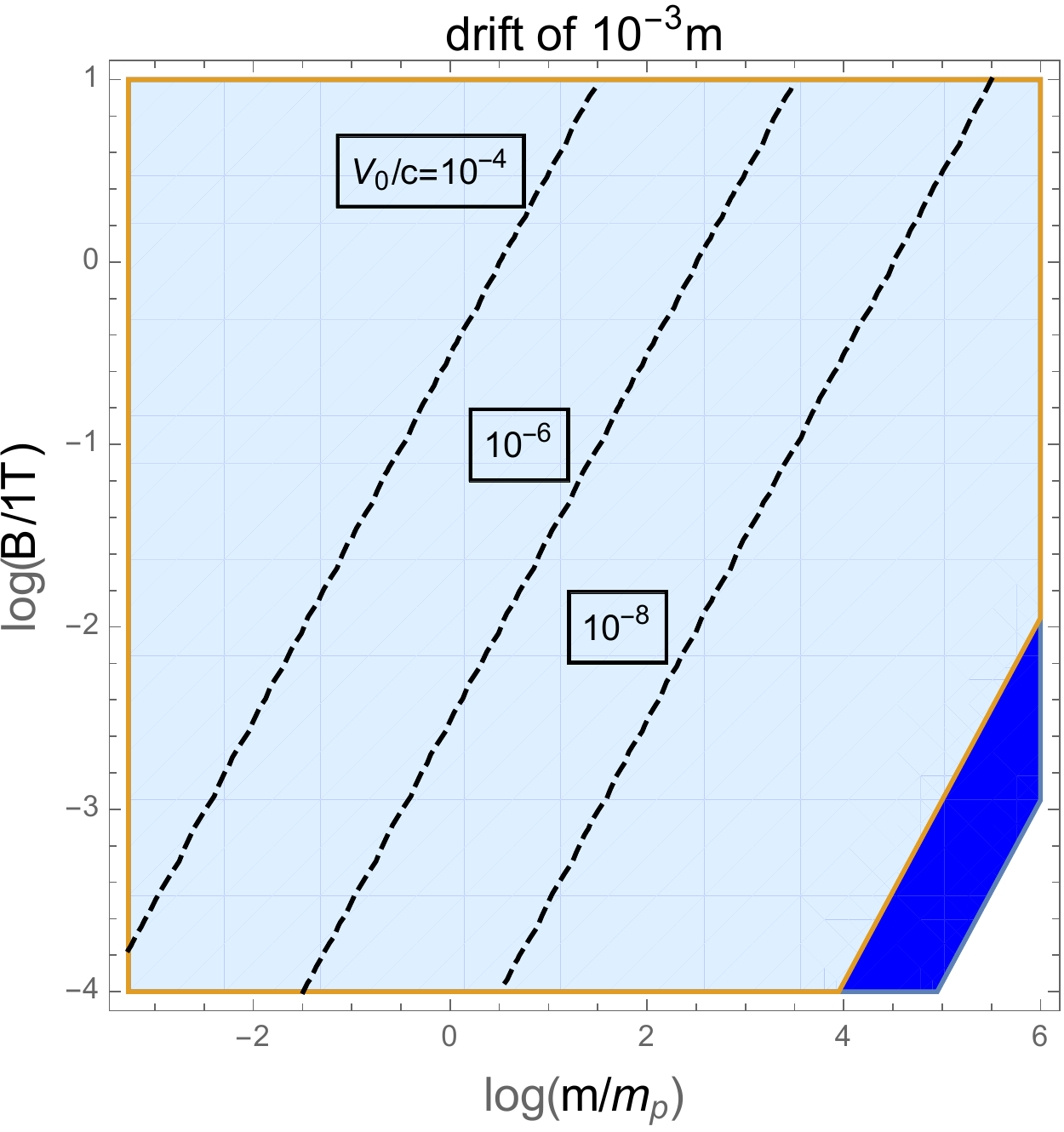}
   \caption{Constraints on the free experimental parameters $(B,m)$ for a particle of charge $Z=1$ and a force of typical magnitude  $F_0~\sim 10^{-7}$~N/kg for a drift of
   $10^{-6}$~m (left) or $10^{-3}$~m (right) over a time scale smaller than 1~hr (white region) or 10~hr (blue region) along the circle of radius $r_c$. The dashed lines indicate the values of $V_0/c$, showing that a non-relativistic description is sufficient.}
   \label{Fig4} 
\end{figure}

The free parameters at hand are $\omega_0$ (fixed by the choice of the particle and the magnetic field) and $V_0$ (fixed by the initial kinetic energy. This defines the radius of the free trajectory. If we start from $(x_0,y_0)=(0.2,0)$ [in meter] with $\alpha=\pi/2$ we need $R_0=0.2$~m for the trajectory to remain inside the two cylinders. Assume that 
\be\label{ansatz}
\varphi=(a/r)
\ee 
so that $F=-\beta a c^2/r^2$ that we normalize to have an amplitude of $F_0=\beta a c^2/r_c^2~\sim 10^{-7}$~N/kg on $r_c=0.4$~m so that $F=-F_0(r_c/r)^2$. It follows that we get
$$
\omega_{\rm drift} = \frac{F_0}{r\omega_0}.
$$
This shows that the time for the orbit to drift from a distance $R$ is $\tau=R/rc\omega_0$. We have the constraints that $V_0<c$ while we want to optimize the drift. To get some insight let us consider the time for the orbit to drift from a length $R$ and ask whether this could be smaller to a time scale of some hours. This sets the constraints
\be
R_0\omega_0(B,m,Z)<c,
\quad
\tau(R,B,m,Z) =\frac{R\omega_0}{F_0}<T_{\rm exp},
\ee
$T_{\rm exp}$ being the duration of the experiment. The second relation implies that if $R\sim R_0$, i.e. a drift comparable to the gyroradius,  then $R_0\omega_0<3.6\times10^{-4}$~m/s for $F_0~\sim 10^{-7}$~N/kg so that the first constraint will always be satisfied. Using Eq.~(\ref{e.w0}), this implies
\begin{eqnarray}
\!\!\left(\frac{B}{1{\rm T}}\right)\left(\frac{m}{m_p}\right)^{-1}\!\! &<&3.8\times10^{-9} \frac{(T_{\rm exp}/1~{\rm h})}{Z}\nonumber\\
&&\quad\times  \left(\frac{F_0}{10^{-7}{\rm N/kg}}\right)\!\left(\frac{R}{10^{-3}~{\rm m}}\right)^{-1}
\end{eqnarray}
which gives the constraint on $(B,m)$ that would allow one to observe a drift of $R$ on a time scale of $T_{\rm exp}$. As can be read from Fig.~\ref{Fig4}, a typical drift of $1~\mu$m on a timescale smaller than 1~hr could be observed for a magnetic field of 1~mT and a particle of $100~m_p$. These orders of magnitude can be recovered from the distance drifted in a time $\tau$ as
\be
\frac{R_{\rm drift}}{1~{\rm cm}} = 3.8\times 10^{-3}\left(\frac{B}{10^{-3}{\rm T}}\right)^{-1}\!\!\left(\frac{m}{100m_p}\right)Z^{-1} \left(\frac{\tau}{1~{\rm hr}}\right).
\ee

\subsubsection{\it Generic chameleon} 

In Ref.~\cite{us2}, we have shown that we can generate a field profile that depends on $\theta$ by shifting the axis of the inner cylinder by $\delta$. The amplitude of the monopoles were shown to be proportional to $\delta/R_{\rm in}$ and to decrease with the multipole. 

The main effect of an angular dependence is that the angular momentum will not be conserved since
\be\label{e.51}
\dot{\cal E}=0,
\qquad
\dot\ell_z = \beta c^2\partial_\theta\varphi.
\ee
Since the angular momentum will vary along the trajectory, it implies that the inner and outer radius of the annulus of allowed trajectories will change over time. Indeed, it is still given as the root of Eq.~(\ref{e.annulus}) with $\dot r$ given by Eq.~(\ref{e.cons01}) but $\ell_z$ is no more constant.

\begin{figure}[htb]
\centering
 \includegraphics[scale= 0.3]{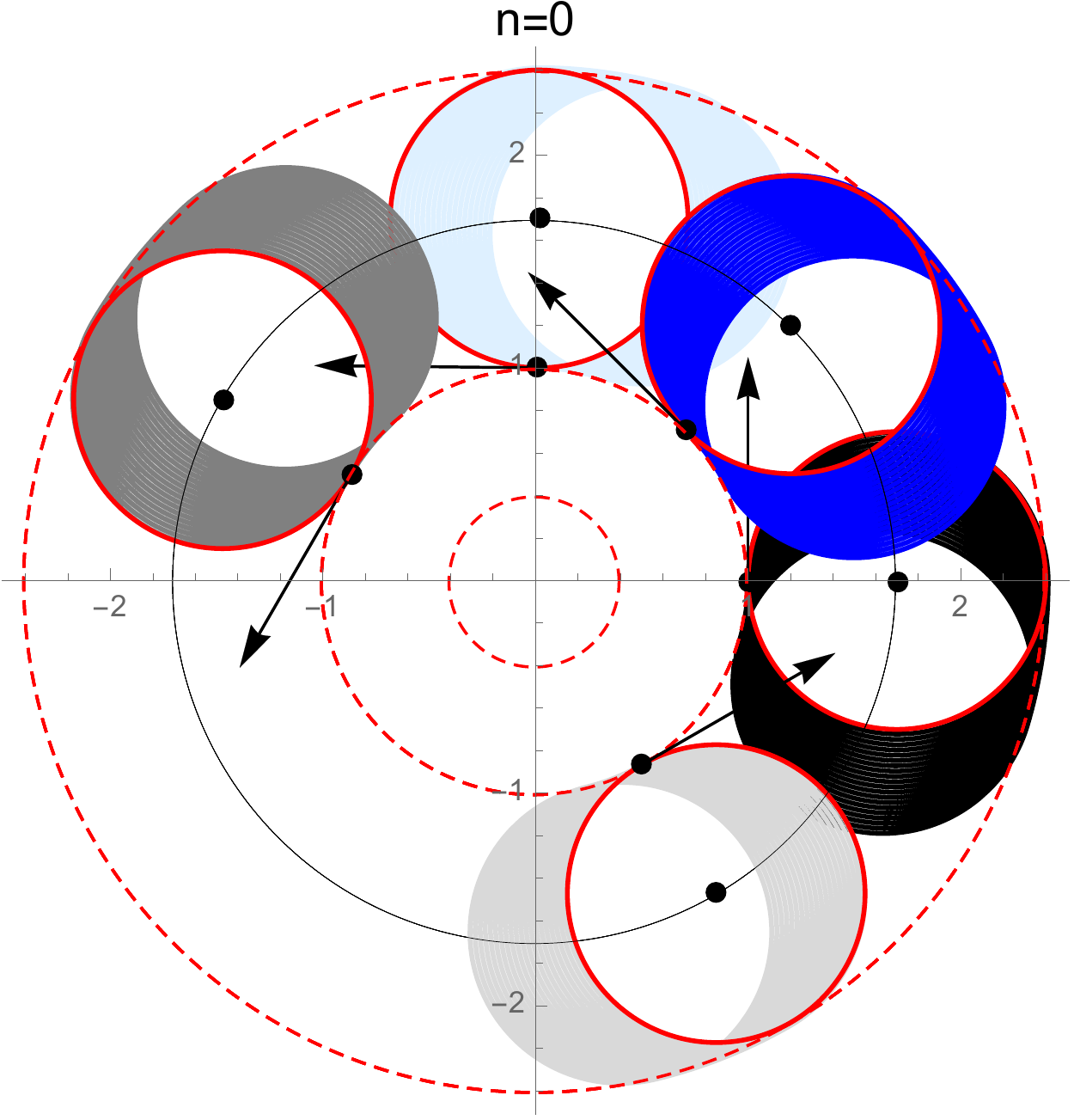} \includegraphics[scale= 0.3]{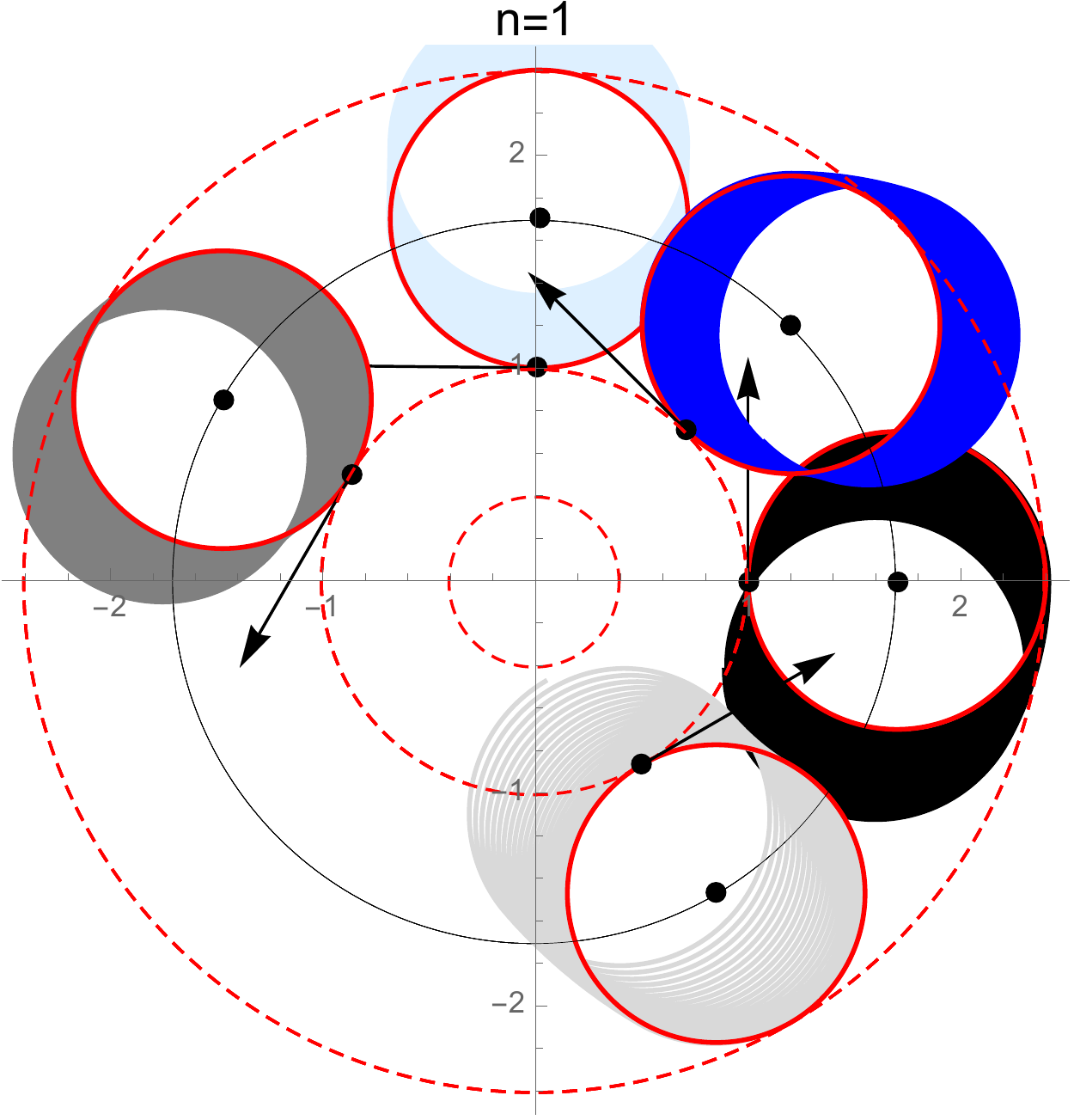}
  \includegraphics[scale= 0.3]{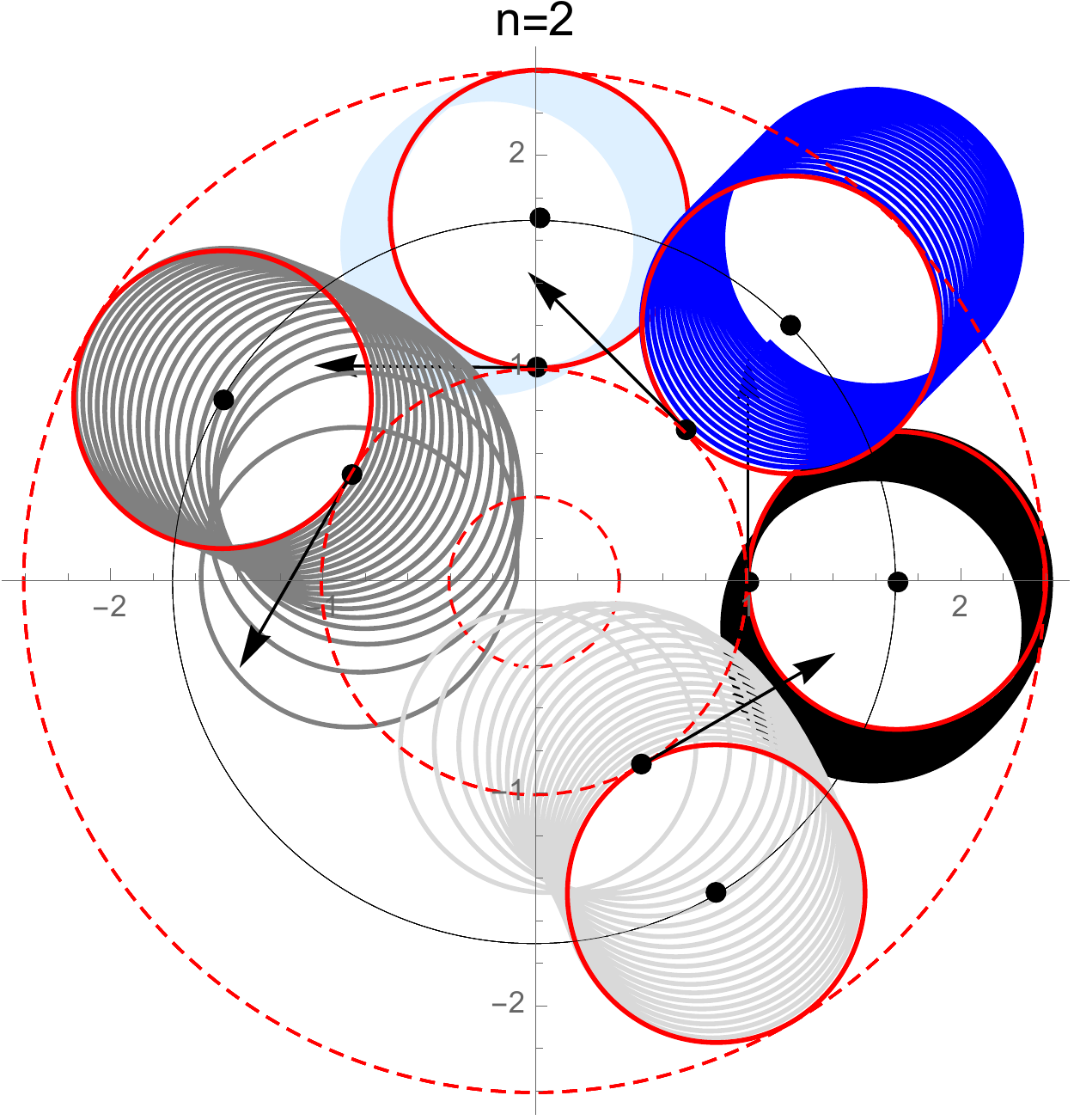} \includegraphics[scale= 0.3]{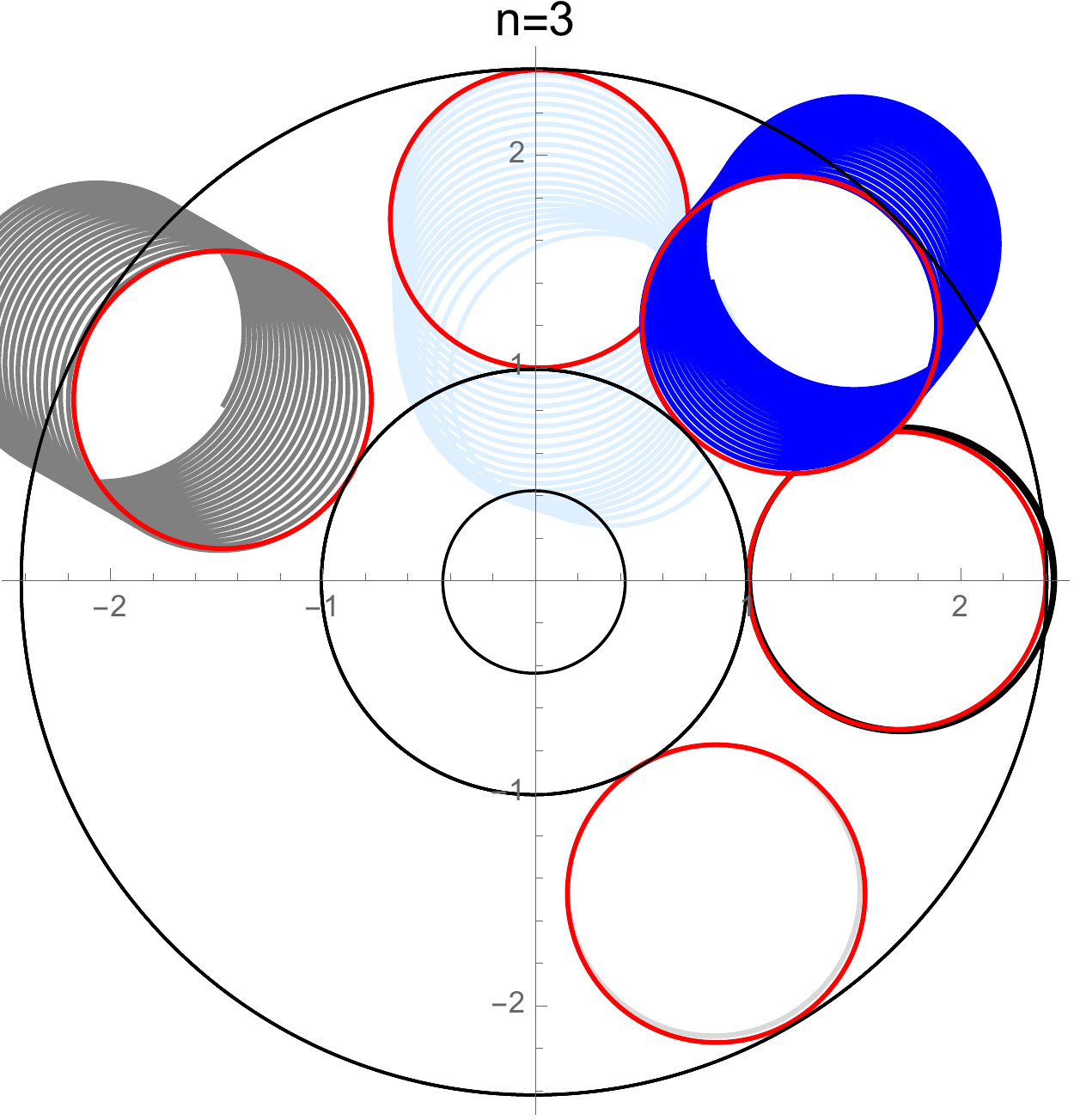}
   \caption{Drift patterns for the 4 first multipoles ($n=0\ldots4$) assuming the form~(\ref{e.potrad}) for the field configuration with $\Phi_n = {a}/{r}$ for all $n$.  All plots assume $\omega_0=1$~s$^{-1}$, $V_0=0.7$~m/s, and $\beta c^2 a=0.01$~m$^3$/s$^{2}$ tangent to the circle with $r=1$~m initially with initial angle $\theta_0=$ 0 (black), $\pi/4$ (blue), $\pi/2$ (light blue), $5\pi/6$ (gray), $5\pi/3$ (light gray) so that the colors represent trajectories with same initial conditions.}
   \label{Fig5} 
\end{figure}

Then, the drift of the trajectory will not be orthoradial anymore as in Eq.~(\ref{e.driftr}). Assume for the sake of the argument that the field configuration is the sum of  multipoles of the form
\be\label{e.potrad}
\varphi_n(r)=\Phi_n(r)\cos n\theta,
\ee
to which one shall add multipoles in $\sin n\theta$, that we omit since it does not modify our general argument. Then the fifth force will be the sum of the multipoles
\be
{\bm F}_n = -\beta\left[\Phi_n'(r)\cos n\theta {\bm e}_r - n \frac{\Phi_n(r)}{r}\sin n\theta {\bm e}_\theta
\right]
\ee
so that the drift velocity is
\be
{\bm v}_{\rm drift}^{(n)} = \frac{\beta}{\omega_0}\left[n \frac{\Phi_n(r)}{r}\sin n\theta {\bm e}_r+\Phi_n'(r)\cos n\theta {\bm e}_\theta
\right].
\ee
If $\Phi_n = {a_n}/{r}$ as assumed in our previous example, then the fifth force induced drift will make an angle $\alpha_n=\arctan[n\tan n\theta]$ with respect to the radial direction so that the radial drift is boosted by a factor $n$ compare to the orthoradial drift. This is illustrated on Fig.~\ref{Fig5}. This opens new ways of testing the fifth force since instead of monitoring the drift, one can monitor the charge of the inner or outer cylinder that will change due to the inward or outward drifts of the particle that otherwise would have remained inside the two cylinders.

\begin{figure}[htb]
\centering
 \includegraphics[scale= 0.3]{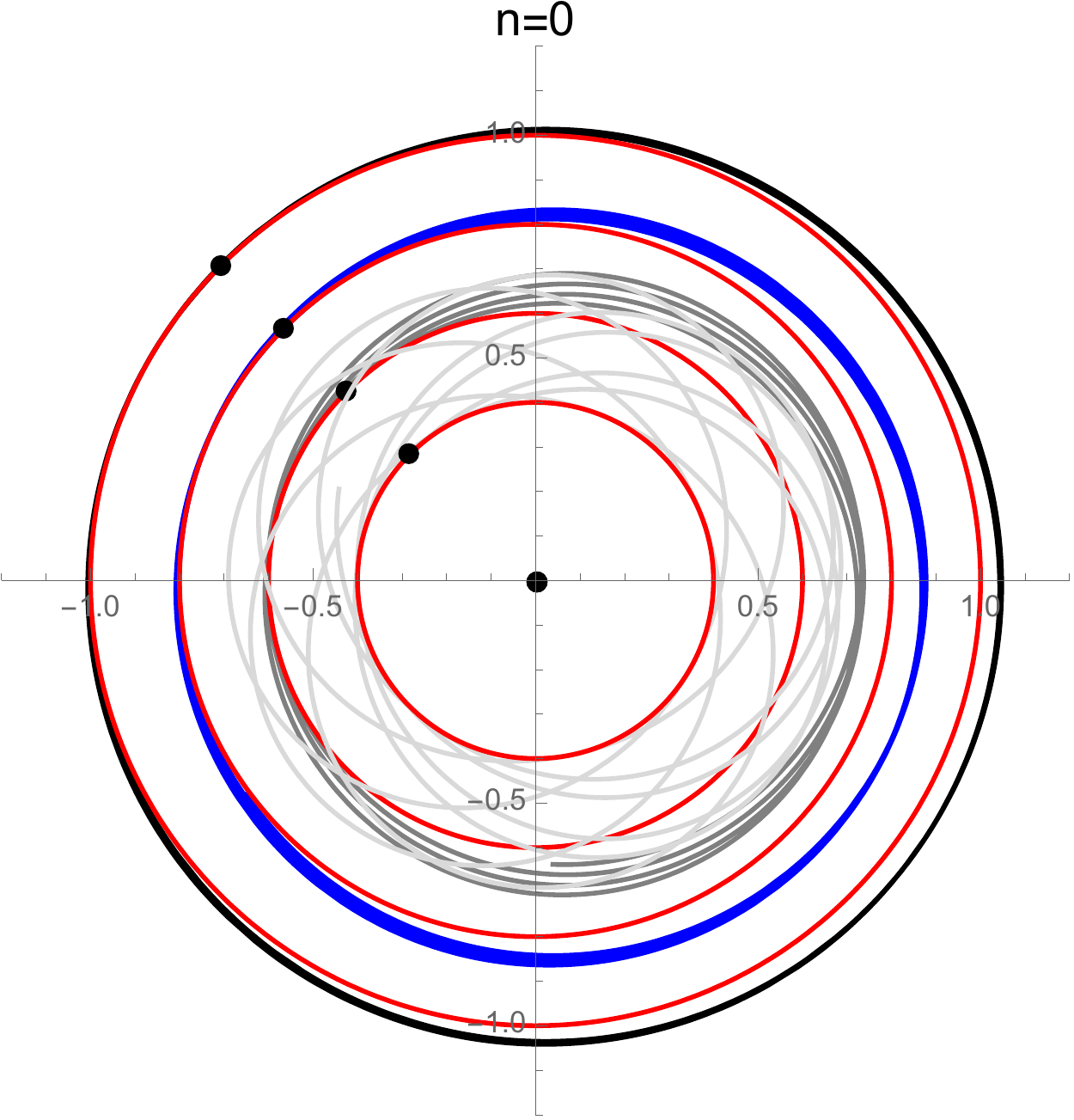} \includegraphics[scale= 0.3]{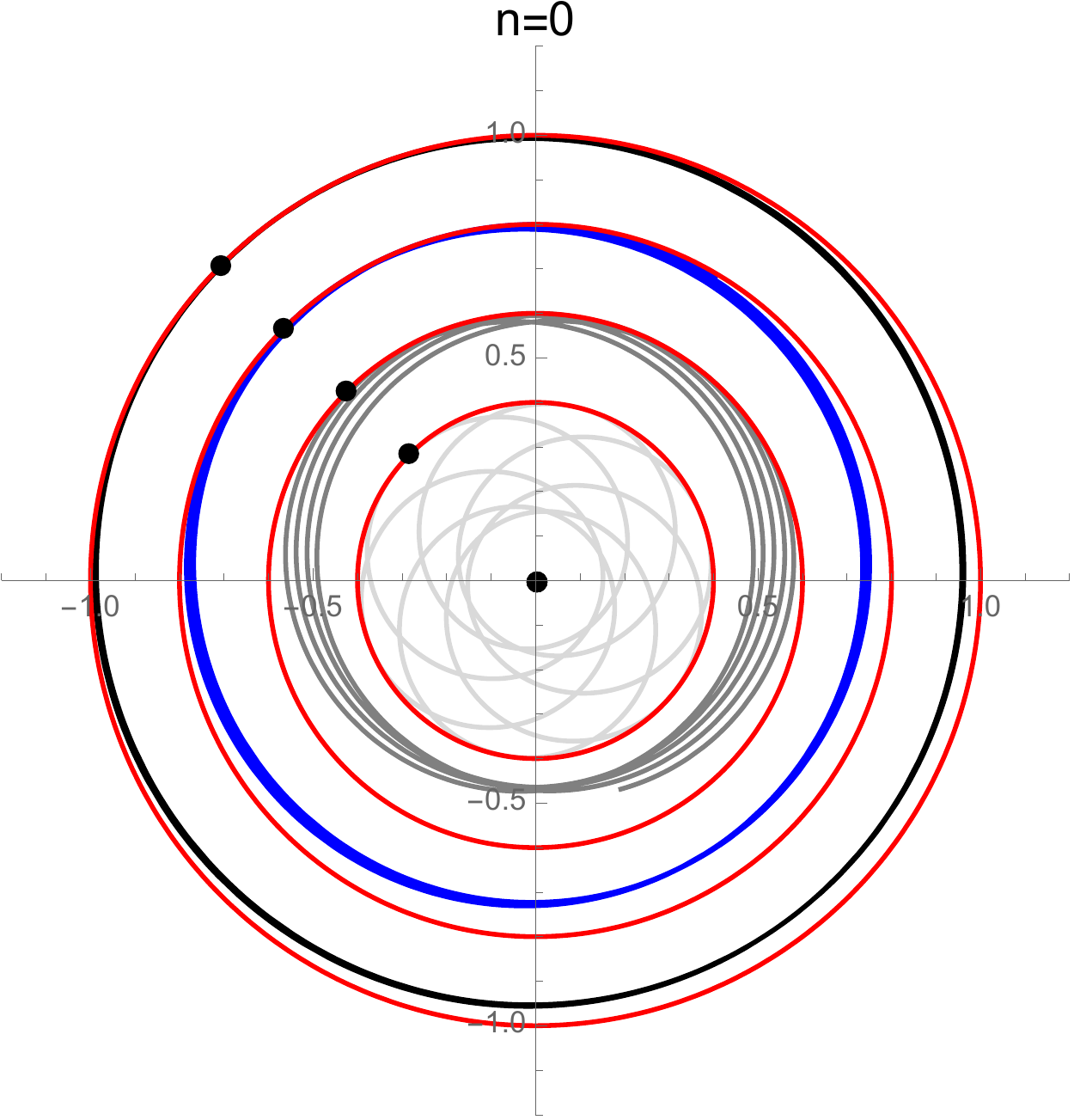}
   \caption{Trajectories tuned such that the gyrocenter coincides with the center of symmetry $O$ initially. In absence of fifth force the trajectory shall be a circle of center $O$. With a fifth force, the trajectory will deviate from this ``free'' trajectory in a couple of gyro-periods. All plots assume $\omega_0=2$~s$^{-1}$, $V_0=0.7$~m/s, and $\beta c^2 a=0.1$~m$^3$/s$^{2}$ [left] and $\beta c^2 a=-0.1$~m$^3$/s$^{2}$ [right] tangent to the circle with $r=1$~m, $0.8$~m, $0.6$~m and $0.4$~m initially with initial angle $\theta_0=\pi/4$.}
   \label{Fig9} 
\end{figure}

To finish, let us also illustrate the effect of the fifth force on trajectories that would be circles of center $O$ in absence of a fifth force. In that case, the guiding center approximation will not hold and the effect of the fifth force can only be investigated numerically. Fig~\ref{Fig9} gives some examples of trajectories for a monopole, comparing an attractive and repulsive force. Indeed it assumes a fifth force with an unreallistically large magnitude for the sake of the illustration. The effect of larger multipoles enlarge the landscape of possible trajectories. The question of the best experimental strategy and the design of the field distribution remain to be discussed.

\section{\bf Macroscopic consequences}\label{section4}

So far we have described the microscopic effects of the fifth force on the dynamics of charged particles. Let us now show that it has a macroscopic side related to the drift current associated with the fifth force.

\subsection{One-dimensional current}

\begin{figure}[htb]
\centering
 \includegraphics[scale= 0.33]{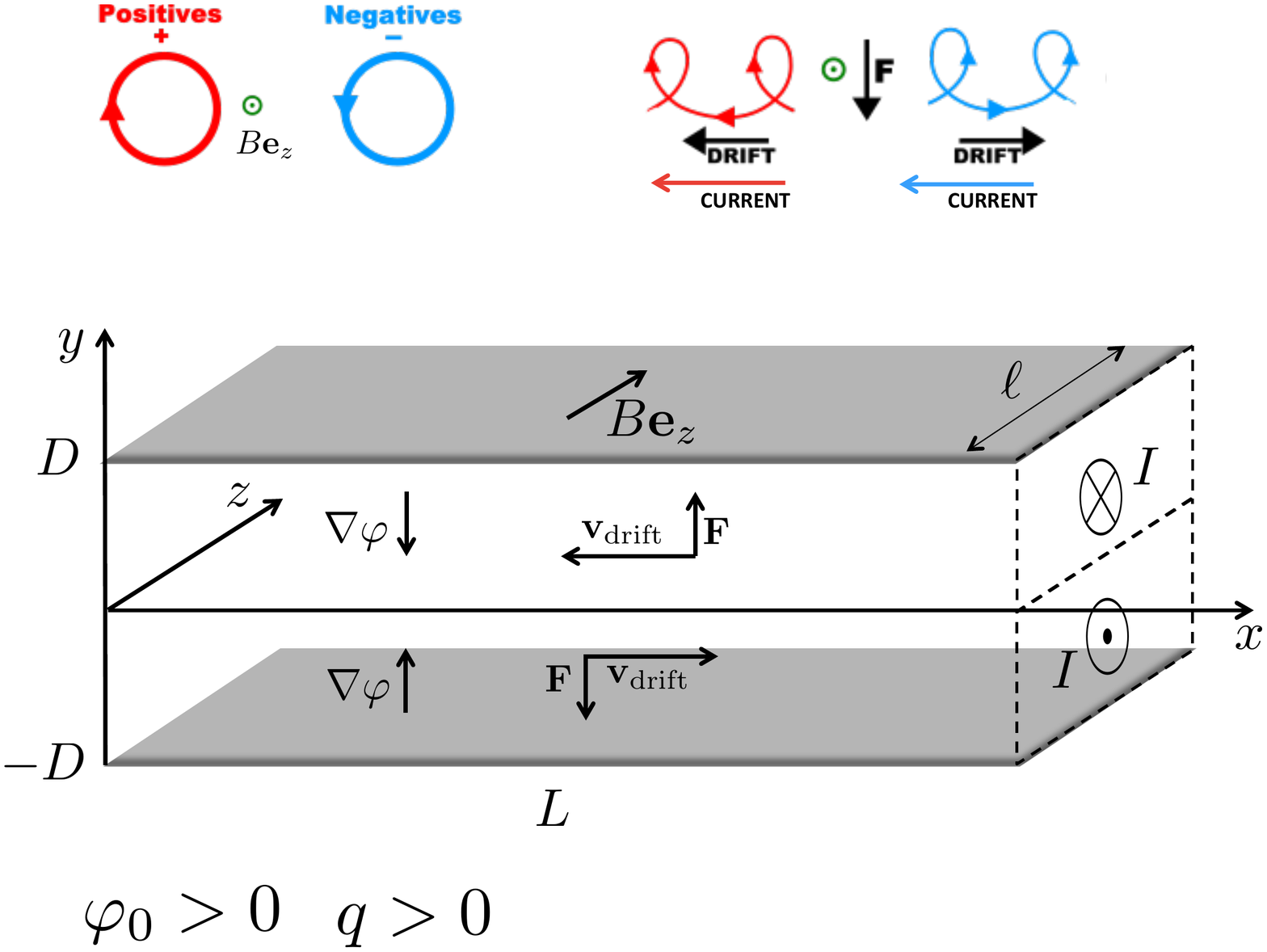}
   \caption{Experimental design to generate macroscopic current from a fifth force. All quantities are defined in the text and are plotted assuming $\varphi_0>0$ and $q>0$. Top pictures show that particles of opposite charges drift in opposite directions but generate a current in the same direction.}
   \label{Fig6} 
\end{figure}

Let us consider two parallel plates as depicted on Fig.~\ref{Fig6} of size $\ell\times L$ along the $xz$-direction and separated by a distance $2D$ along the $y$-axis and assume we impose a magnetic field $B{\bm e}_z$. By symmetry the scalar field will have a profile $\varphi(y)$ so that it generates a fifth force ${\bm F}=-m\beta c^2\partial_y\varphi {\bm e}_y$. 

It follows from Eq.~(\ref{e.drift}) that the particles enjoy a cyclotron motion of pulsation $\omega_0$ drifting along the $x$ axis at the velocity
\be
{\bm v}_{\rm drift}= -\frac{\beta c^2}{\omega_0}{\bm e}_x.
\ee
Now, if the density of charge is $\eta_q $, this generates a current density
\be
{\bm j} = \eta_q q {\bm v}_{\rm drift}(y)
\ee
flowing in opposite directions in the upper ($y>0)$ and lower ($y<0$) parts, because $\partial_y\phi>0$ for $y<0$ and $\partial\phi<0$ for $y>0$. It follows that it will generate a total current 
\be
I = \ell\int_{-D}^D {\bm j}(y).{\bm e}_x \dd y.
\ee

In order to put numbers, let us assume that the profile of $\varphi$ is given by
\be
\varphi=\varphi_0\left(1-\frac{y^2}{D^2}\right)
\ee
so that the force is
$$
{\bm F} =2m\frac{\beta c^2\varphi_0}{D}\frac{y}{D}{\bf e}_y
$$
and we set $F_0=2\beta c^2\varphi_0/D\sim 10^{-7}$~N/kg. Hence, the current density is
\be
{\bm j}(y) = - 2q \eta_q  \frac{\beta c^2\varphi_0}{\omega_0 D}\frac{y}{D}   {\bm e}_x 
\ee
so that the current profile is
\be\label{Iprofile}
\frac{\dd I}{\dd y}=  \ell j(y)
\ee
and the total current
$$
{\bm I} = \mp q \eta_q \ell D \frac{\beta c^2\varphi_0}{\omega_0 D}{\bm e}_x
$$
in the upper/outer region respectively (if $q>0$ and $\varphi_0>0$).
\begin{widetext}
In order to estimate its amplitude, we need assume a typical value of the density. Assume we have a gas in standard conditions, its density is 1~mol/20~l, i.e. 
$$
\eta_q =\eta_0 \sim 3\times10^{25}~{\rm m}^{-3},
$$
then 
\be\label{e.I1}
\frac{I}{1{\rm nA}} = 5\left(\frac{\eta_q }{\eta_0} \right)\left(\frac{B}{1 {\rm T}} \right)^{-1}\left(\frac{m}{m_p} \right)\left(\frac{S_\perp}{ 1{\rm m}^2} \right)\left(\frac{F_0}{10^{-7} {\rm m/s}^2} \right)
\ee
with $S_\perp=\ell D$. First we note that the current is independent of the charge of the particle, simplify because $q v_{\rm drift}$ is, and proportional to the mass. The current reaches  $0.5~\mu$A for $m=m_p$ and $B=0.01$~T.
\end{widetext}

\subsection{Effect on the field profile}

Still, we need to be careful. In the microscopic analysis performed in \S~\ref{section3}, we studied the effect of the fifth force on a test particle and the density inside the cavity was fixed externally. Now, we need to have a large number of particles, with a number density $\eta_0$ so that the mass density inside the cavity $\rho\sim5\times10^{-2}(m/m_p)$~kg.m$^{-3}$. As a consequence  this will affect the profile inside the cavity since the Klein-Gordon equation is
\be
\partial_y\phi(y)=n\Lambda^{n+4}\left[\phi_*^{-(n+1)}-\phi^{-(n+1)}\right]
\ee
in one-dimension, with
$$
\phi_*^{n+1}=\frac{M_{\rm P}\Lambda^{n+4}n}{\beta\rho}.
$$
The field tends toward $\phi_*(\rho_{\rm mat})$ in the wall on a length scale of the order of the Compton length
$$
\lambda=\sqrt{\frac{\phi_*^{n+2}}{n(n+1)\Lambda^{n+4}}}
$$
of the order of $\lambda_{\rm wall}=2$~cm. This shows that one will need to properly design the parameters of the experiment since one would want to increase $m$ and $\eta_q $ to get a higher current, but that would increase the density $\rho_{\rm in}$ so that $\phi_*\propto 1/\rho^{n+1}$ will decrease as well as $\lambda\propto1/\rho^{1+n/2}$ so that the force will scale as
$$
F\propto\partial_y\phi\propto\frac{\phi_*}{\lambda}\propto\rho_{\rm in}^{-\frac{n+2}{2(n+1)}}.
$$
Hence one can either adopt a model-independent approach and constrain $F_0$ for a chosen set $(B,m,\eta_q)$ or one can try to constrain a given  model, in which case the scaling above and the dependence of the force on the density of matter inside could be used to optimize the choice of $(\eta_q ,m)$ since it sets the amplitudes of the current but also affects $F_0$ through the mass density. 
\begin{widetext}
As an example, we provide the profile of the scalar field from which one can deduce the profile of the force and of the current density. These are depicted on Fig.~\ref{Fig8}.
\begin{figure}[htb]
\centering
 \includegraphics[scale= 0.8]{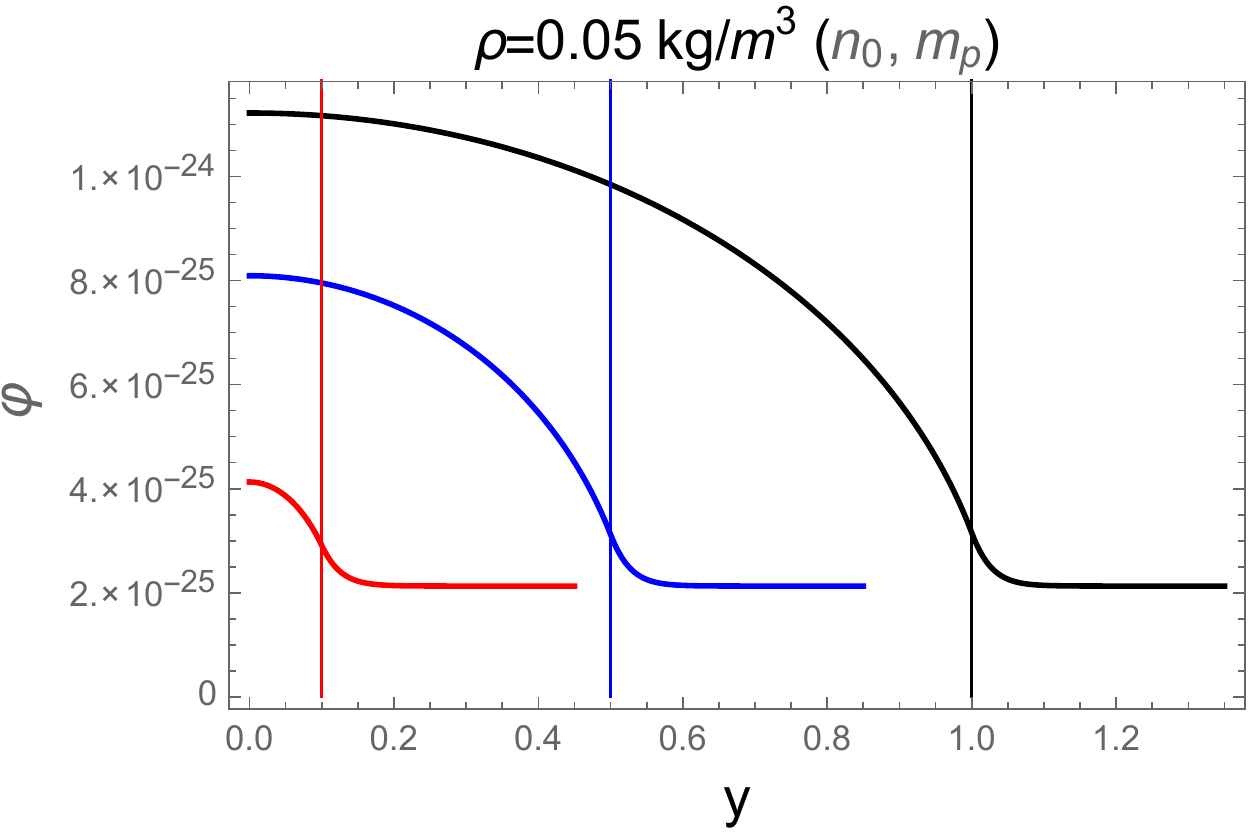}
 
  \includegraphics[scale= 0.45]{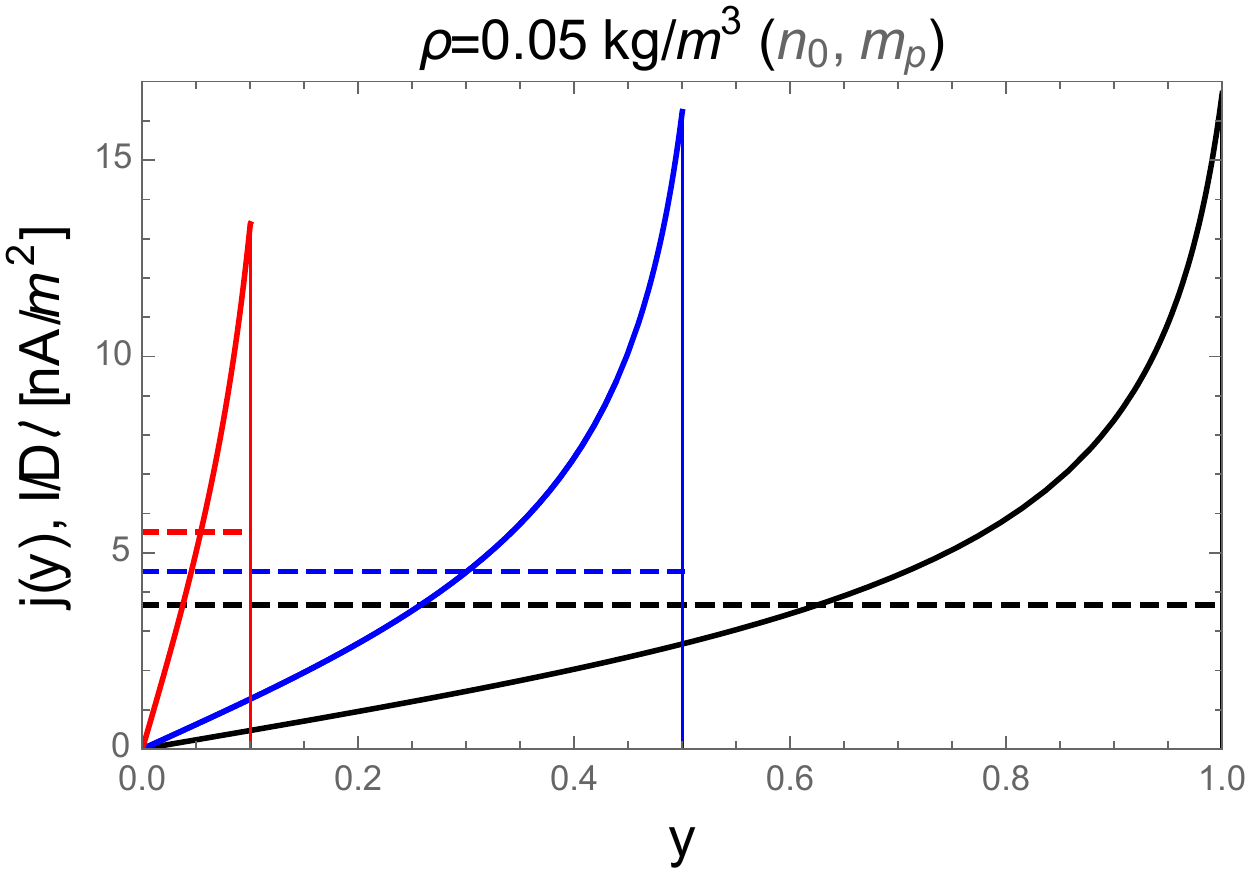} \includegraphics[scale= 0.45]{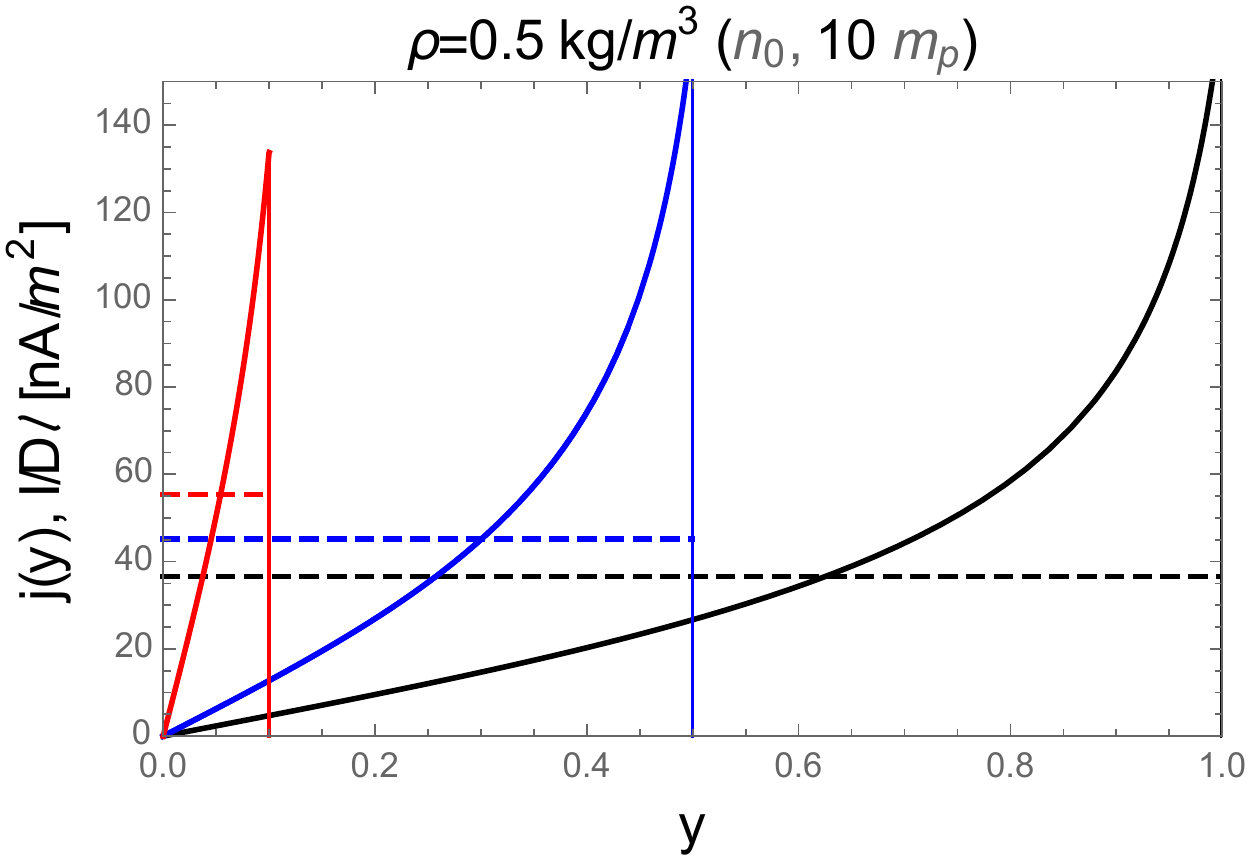} \includegraphics[scale= 0.45]{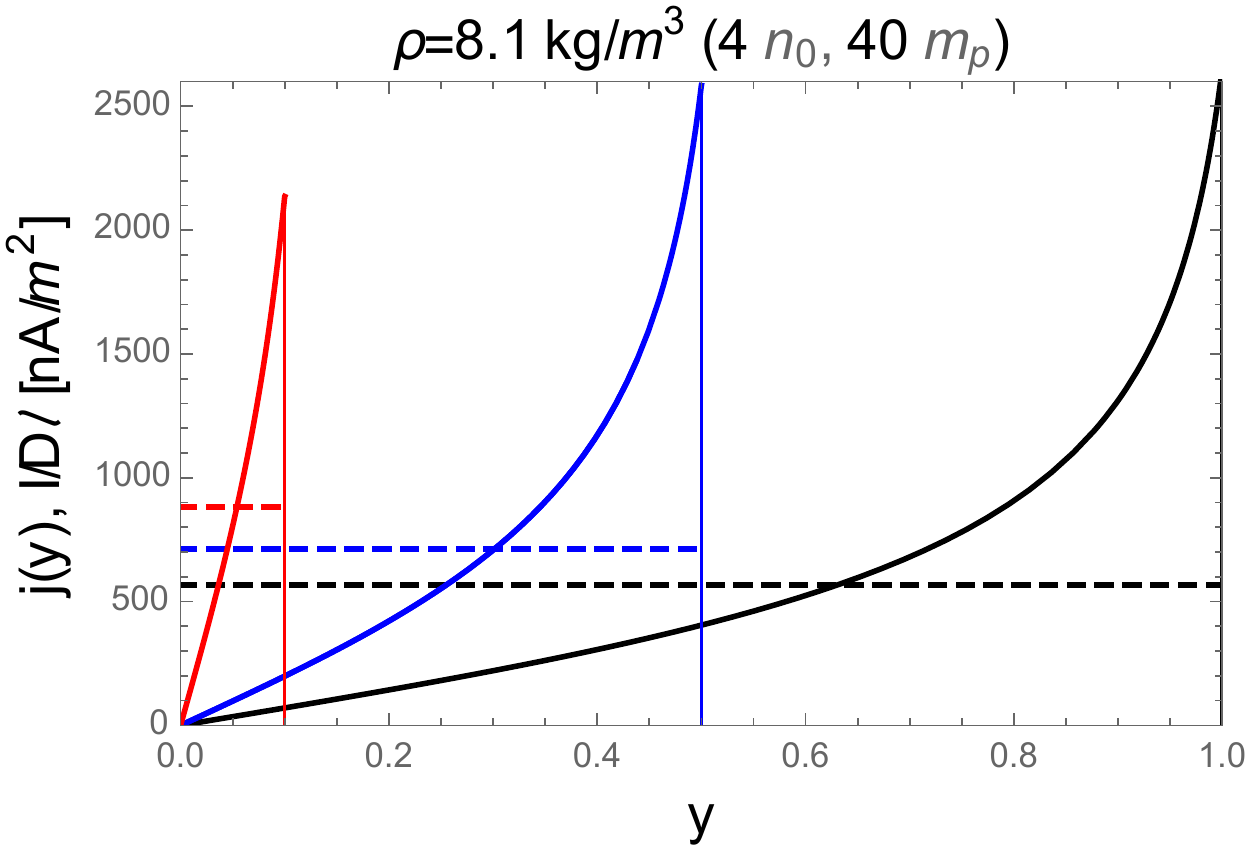}
   \caption{Profile of the scalar field $\varphi(y)$ for a chameleon model with $n=2$, $\Lambda=1$~eV and $\beta=1$ assuming that $D=1,0.5,0.1$~m (Black, Blue, Red) and that the density inside the cavity is $\rho_0=\eta_0 m_p=0.05$~kg/m$^{3}$ (top); the changes in the profile for $10\rho_0$ and $10^{-3}\rho_{\rm mat}$ are not visible by eye. For the same models, we obtain the profile of the current density $j(y)$ (solid lines)  and the total intensity per unit surface (dashed lines), both in nA/m$^2$.}
   \label{Fig8} 
\end{figure}

\subsection{Annular current inside the cylinders} 

Coming back to the case of the nested cylinders we studied earlier, the same reasoning shows that there shall exist an annular current along ${\bm e}_\theta$ given by
\be
{\bm j}(r)=\eta_q q \frac{\beta c^2}{\omega_0}\varphi'(r) {\bm e}_\theta,
\ee
corresponding to a total current
\be
{\bm I}=\eta_qqL \frac{\beta c^2}{\omega_0}\int_{R_{\rm in}}^{R_{\rm ext}}\varphi'(r)\dd r {\bm e}_\theta.
\ee
\end{widetext}
if $L$ is the length of the cylinders. And, as expected from the Lenz law, it generates a magnetic field along the $z$-axis, with typical magnitude on the axis
\be
B_{\rm drift} = \mu_0 \eta_q q \frac{\beta c^2}{\omega_0}\int_{R_{\rm in}}^{R_{\rm ext}}\varphi'(r)\dd r .
\ee
With the ansatz~(\ref{ansatz}) we get the typical magnitudes
\begin{eqnarray}
{\bm I}&=& -\frac{\eta_q q}{\omega_0} F_0 (R_{\rm ext}-R_{\rm in})L{\bm e}_\theta\\
{\bm B}_{\rm drift} &=& \mu_0{\bm I}/L 
\end{eqnarray}
with the permeability of vacuum $\mu_0=4\pi\times10^{-7}$~T.m/A and, again $F_0=\beta ac^2/R_{\rm ext}R_{\rm in}$. The typical order of magnitude is identical to the one of Eq.~(\ref{e.I1}) with $S_\perp=2L(R_{\rm ext}-R_{\rm in})$. It can then be checked that $B_{\rm drift}\sim 10^{-18}$~T so that it can be completely neglected compared to the experimental magnetic field.

\subsection{Discussion}

This shows that the effect of the fifth force on the dynamics of a charged particle at the microscopic level has several macroscopic consequences: (1) in 1 dimension, it generates a drift current between the parallel walls, (2) in 2 dimensions with cylindrical symmetry, it generates an annular current and (3) in 2 dimensions with no cylindrical symmetry, the particles drift inward and/or outward and may charge the walls of the cylinders, leading to the growth of a radial electric field.

Our numerical estimations~(\ref{e.I1}) favor high mass particles, with no dependencies on its charge, while at the microscopic level, the drift effect favors large mass, low charge particles. A key issue is the {\em density} that can be reached in laboratory experiments. Plasma densities typically ranges from $10^3$ to $10^{33}$~m$^{-3}$ in nature. Pushing to $10^{20}$~m$^{-3}$ will allow one to get a current larger than 1~nA. Note also that in the one-dimensional set-up, one can in principle access $I(y)$. Such a measurement would be extremely valuable since it will enable to get some information on the profile $\varphi(y)$, i.e. it potentially gives access to a way to constrain the parameters of the model -- see  Eq.~(\ref{Iprofile}).

Note also that the {\em temperature} of the plasma is not a key issue since the drift is insensitive to the velocity of the particle. Nevertheless, we need it to be cold enough so that the gyroradius is much smaller than the typical size of the experiment, i.e. we shall demand that $\sqrt{2k_BT/m}/\omega_0^2\ll 1$~m, i.e. that $T<10^{11}$~K, which is achieved easily for protons.

To finish let us remind  that there is a force much larger than the fifth force that causes the particle to drift: the {\em standard gravitation} since its magnitude is of order $g=10$~m/s$^{2} $ and thus would cause a drift typically 9 orders of magnitude larger, at least, than the one induced by the fifth force. Luckily we can suppress this effect: since the drift~(\ref{e.drift}) behaves as ${\bm F}\wedge {\bm B}$, aligning the magnetic field with the local gravitational field will ensure that it will not act on the particle. This can be done in a table-top experiment for a chameleon field since its profile is dictated by the geometry of the experiment and screened from the local environment. Actually, it offers a nice way to calibrate the experiment. Since $g\gg F_0$ one can first set the walls vertical so that the magnetic field is horizontal and measure the current $I_{\rm max}$  and then rotate the whole experiment until the magnetic field is vertical. Hence the current shall change as
$$
I=I_{\rm max}\left[\sin\theta + \frac{F_0}{g}\right].
$$
The measurement of $I_{\rm max}$  and of the local gravity field allows one to evade the individual measurement of $\eta_q$ and $B$. Then, any upper limit on $I(0)$ provides a constraint on $F_0/g$. Concerning the Newton force induced by the walls of the cavity, first let us remind that it will strictly vanish if the walls are infinite. Then, for large parallel walls, the residual gravity field has a component parallel to the magnetic field; it induces no drift while only its $y$-component has an effect that will modify the total current while the $x$-component will modify the profile of the current density. The amplitude of $g_y$ is smaller than $G\rho_{\rm mat} e D/L\sim 5\times10^{-8}(e/10~{\rm cm})(D/L)$~m/s$^2$ hence roughly 2 orders of magnitude smaller than the fifth force we try to measure. Hence to maximise the current, we need to maximise the surface, i.e. $\ell D$, while minimizing $D/L$ in order to make the gravity of the walls completely negligible. As can be shown from Fig.~\ref{Fig8} it also gives a higher mean current density.

Let us also stress that in the discussions of \S~\ref{section3} we have not included the effect of the Newtonian gravitational field induced by the cylinders. First, if the cylinders are infinite the Newton force in the inter-cylinder space vanishes exactly. Then for finite length cylinders, for the radial set-up, the gravitational force will be aligned with the axis of the cylinders, and thus with the magnetic field so that it will induce no drift. When the cylinders are not coaxial, there will be a small residual Newton force that will be, similarly to the case discuss in the previous paragraph, negligible.

To finish, let us mention a possible way to increase the sensitivity. As seen from Eq.~(\ref{e.51}) the angular profile of the force affects the evolution of the angular momentum which is not conserved anymore when there is no cylindrical symmetry. One can think of designing the shape of the inner and outer ``cylinders" so that the profile exhibits sharp changes in $\partial_\theta\varphi$, similar to electric point effect. That could generate locally large gradients, the design of which could be controlled and hence distinguished from other forces. Such ideas need to be investigated later.

All these arguments convince us that this can provide a new experimental concept to detect fifth force in the laboratory. Indeed for now we just established orders of magnitude for such an experimental set-up, the technological feasability of which would need to be investigated in details, a task much beyond the scope of this work.

\section{\bf Radiation damping} \label{section5}

Besides the fifth force and the magnetic force, the particle being accelerated shall undergo a reaction force, the Abraham-Lorentz-Dirac force, the effect of which needs to be compared to the fifth force.The equations of motion  have to be extended to
\be
m\gamma^\mu = q\left(F_{\rm ext}^{\mu\nu} + F_{\rm self}^{\mu\nu}\right)u_\nu,
\ee
in Gaussian units, where $F_{\rm ext}^{\mu\nu}$ is the Faraday tensor of the electromagnetic field of the moving charge. The computation of the reaction forces requires to evaluate the self-retarded potential. This is detailed in chapter II.19 of Ref.~\cite{DUbook}. It requires a regularization and many schemes are used in electrodynamics, see e.g. Ref.~\cite{damour75}.  Using a regularization by averaging on the direction gives the radiation reaction force
\be
F_{\rm self}^{\mu\nu}u_\nu= \frac{2}{3}q\left(\dot\gamma^\mu -\gamma^2 u^\mu\right)
\ee
as proposed by Abraham, Lorentz and Dirac. In the non-relativistic limit, the radiation reaction force takes the form
\be
{\bm F}_{\rm reac}= \frac{2}{3}\frac{q^2}{4\pi\varepsilon_0 c^3}\dot{\bm a},
\ee
once we put the international units back.

It is easily evaluated on the free trajectory since ${\bf V}=V_0(\cos\omega t-\alpha,\sin\omega t)$. It is indeed a damping force
$$
{\bm F}_{\rm reac}=-\frac{2}{3}\frac{q^2}{4\pi\varepsilon_0 c^3}\omega_0^2{\bm V}.
$$
This implies that it does not induce a drift but a shrinking of the trajectory so that it cannot be confused with the effect of the fifth force. Nevertheless, it needs to be evaluated since it will limit the duration of the experiment.

\section{\bf Conclusion}

This article has investigated the effect of a small fifth force of scalar origin on the dynamics of a charged particle. It has derived the full relativistic equations of motion and conserved quantities and gave their non-relativist limit. Then, it investigated the dynamics of a charge in a uniform magnetic field to show that the standard cyclotron motion enjoys a drift, similar to the one that can be observed if the magnetic field is not uniform. This drift is fully dictated by the profile of the scalar field. Focusing on profiles in between two nested cylinders, as studied in our previous works~\cite{PhysRevD.100.084006,us2}, we have shown that the drift is orthoradial if the configuration is cylindrically symmetric and has a more involved angular structure for a general profile.

One can control the cyclotron pulsation $\omega_0$ by choosing the particle and tuning the magnetic field. Controlling the initial velocity of the particle determines its gyroradius. Then, the typical  properties of the drift (timescale and direction) depend on the fifth force, that is on the profile of the scalar field within the two cylinders. While the profile of a light scalar field cannot be tuned for a light dilaton, this is not the case for a chameleon field. Thanks to the environmental dependance, the field inside the cavity is screened from the outside and its profile will mostly depend on the local density in the cavity, the nature of the walls and the geometry of the cavity. This is a crucial property of these models, allowing one to engineer these fields (indeed if they exist). In particular, and as demonstrated in Refs.~\cite{PhysRevD.100.084006,us2}, shifting the axis of the cylinders allows one to design angular dependencies. The typical amplitude and profile of the force will depend on the parameters of the microscopic model $(\Lambda,n,\beta)$ and the design of the experiment $(R_{\rm in},R_{\rm ext},\delta,\rho)$ and was shown to be typically of the order of $10^{-7}$~m/s$^2$. We already mentioned in Ref.~\cite{PhysRevD.100.084006} that the force affects any experiment based on monitoring the trajectory of atoms inside a cylindrical cavity of free falling particles in space. 

These effects on individual particles would require to monitor a drift, or relative drift, of single particles on the order of the gyroradius on a time scale of the hour for a force of $10^{-7}$~m/s$^2$. As explained, there is a macroscopic side to these effects since the fifth force induces macroscopic currents that may be easier to measure. In that case we need to have a plasma within the cavity, which would affect the force and its profile since it modifies the local mass density inside the cavity. In the particular case of the one-dimensional experimental set-up proposed in this work shows that a fifth force of $10^{-7}$~m/s$^2$ can induce a drift current drift larger than 5~nA. This would require to push the density to  the density of a gas in standard conditions while the density of plasma in nature can range from $10^3$ to $10^{33}$~m$^{-3}$. Hence the density is one of the key parameters. Otherwise one would need to operate with a magnetic field of $1~\mu$T and heavy particles. The temperature of the plasma plays no major role since the drift velocity is independent of the energy of the particle. Nevertheless we shall require that the gyroradius is much smaller than the typical size of the experiment. Setting $R_0\sim\sqrt{2 k_BT/m}/\omega_0\ll 10^{-3}$~m implies that the temperature be smaller than $5\times10^{6}$~K, which is easily achieved -- room temperature would correspond to $R_0\sim20~\mu$m. It is also important to remind that the effect of gravitation, that also induces a drift several orders of magnitude larger, can be screened by aligning the magnetic field with the local gravity field. As a consequence, it is not necessary to go to space. Then, the gravity of the walls of the cavity are roughly 2 orders of magnitude smaller than the nominal fifth force we could measure. Given these numbers, the feasibility or the existence of loopholes in our arguments would require to be investigated with care. Note also that the experiment may also enable to access the transverse profile of the chameleon field, directly related to the properties of the potential and coupling function, a possibility which has not been offered by any other proposed experimental set-up so far.

Indeed, it would be bold to argue that it offers so far a new experimental design to test fifth force in laboratory. We have just used toy field profile to illustrate the physical effects and derive orders of magnitude. One would need to implement, and most probably optimize, field profiles, as shown in Ref.~\cite{us2} and discuss the detectability of the drifts and of the current and all sources of noise that will unavoidably be present. The question of the alignment of the magnetic field with the local gravity field is crucial as well as a careful study of the gravity induced by the surrounding of the experiment. To finish, we note that we still have the freedom fo let the magnetic field vary in time. 

Nevertheless we believe that it opens a way of reflection to eventually reach such a new experimental set-up. Let us also  mention, to finish,  that the equations of motion derived here are fully general and can also be applied to the propagation of cosmic rays.

\acknowledgements{We thank Philippe Brax, Gilles Esposito-Far\`ese, Pierre Fleury, Julien Larena, Roland Lehoucq, Cyril Pitrou, and Manuel Rodrigues for their comments and insight.}

\bibliographystyle{ieeetr}
\bibliography{Bib}

\appendix

\section{Initial conditions}

The initial conditions can be fixed by either choosing $({\bm x}_0,V_0,\alpha)$ or $({\bm x}_0,{\cal E},\ell_z)$. The first are more natural since one does not know the potential $\varphi$ but the second allows one to compare motion with the same constants of motions.

One can easily shift from one to the other since
\begin{itemize}
\item Starting from  $({\bm x}_0,V_0,\alpha)$, we have $r_0=\sqrt{x_0^2+y_0^2}$, $V_{x0}=V_0\cos\alpha$, $V_{y0}=V_0\sin\alpha$, ${\cal E}=V_0^2/2+\beta\varphi(r_0)$, $\theta_0=\arctan(y_0/w_0)$ so that $\dot\theta_0=V_0\sin(\alpha-\theta_0)/r_0$ and then $\ell_z=r_0^2(\dot\theta_0+\omega_0/2)$.
\item Starting from$({\bm x}_0,{\cal E},\ell_z)$, we have $r_0=\sqrt{x_0^2+y_0^2}$ so that $V_0=\sqrt{2({\cal E}-\beta\varphi(r_0))}$. Then, $\theta_0=\arctan(y_0/w_0)$ and $\dot\theta_0= (\ell_z/r_0^2 - \omega_0/2)$, $V_{\theta0}=r_0\dot\theta_0$ so that $\alpha= \theta_0+\arcsin(V_{\theta0}/V_0)$ and then $V_{x0}=V_0\cos\alpha$, $V_{y0}=V_0\sin\alpha$.
\end{itemize}

It is also interesting to rewrite the dynamical system by using the dimensionless time $\tau=\omega_0 t$ and rescaling the lengths in units of the gyroradius $R_0$ as
\begin{eqnarray}
x''&=&y'-\gamma\frac{x}{r^3},\nonumber\\
y''&=&-x'-\gamma\frac{y}{r^3},
\end{eqnarray}
with the dimensionless parameter $\gamma =\beta c^2 a/\omega_0^2R_0^3$ if the field configuration is given by $\varphi=a/r$. The initial conditions are then given by $v_0=1$ so that $(x'_0,y'_0=(\cos\alpha,\sin\alpha)$ and $(x_0,y_0)$. Under such a form, the dimensional analysis implies that the drift pulsation can only be a function of $(\gamma,r_c)$.

It is easily check that for $\gamma=0$ we have a circular orbit, that is drifting when $\gamma\ll1$ and tend to a precessing ellipse for large $\gamma$ and a standard static ellipse for $\gamma=+\infty$.
 
\section{Particle in an electric field}

For the sake of completeness, let us  consider the case of a one-dimensional electric field between two plates, ${\bm E}=E{\bm e}_x$ so that the only non-zero component of the Faraday tensor is $F^{0x}=E$.

\subsection{Standard acceleration}

When the fifth force vanishes, it is clear from the equation of motion~(\ref{e.motion}) that the 4-acceleration has a constant modulus
\be
\gamma_\mu \gamma^\mu = \left(\frac{qE}{m}\right)^2\equiv g^2.
\ee
This is indeed easy to understand since in the inertial frame tangent to the charge worldline, the electric field remains unchanged in a Lorentz transformation. It follows that $\dd U^x/\dd\tau = g U^0$, i.e. $\dd^2 X/\dd\tau^2=g\dd T/\dd\tau$ with the constraints $U_\mu U^\mu=-(\dd T/\dd\tau)^2+(\dd X/\dd\tau)^2=-1$. It can be integrated as
\be
gT=\sinh g\tau,\qquad
gX=\cosh g\tau
\ee
giving the trajectory
\be
gX=\sqrt{1+g^2T^2}.
\ee

\subsection{Effect of the fifth force}

We now consider that two parallel infinite plates so that the field configuration in between them is given by $\phi(x)$. Indeed since the force is extremely weak, typically smaller than $10^{-7}$N/kg, see e.g. Ref.~\cite{PhysRevD.100.084006}, it will always be subdominant. Nevertheless, it has been argued that such a small force  may affect any experiment based on
monitoring the trajectory of atoms inside a cavity~\cite{Llinares:2018mzl}. Indeed the force has to be compared to gravity and it has been pointed out in Ref.~\cite{PhysRevD.100.084006} that in space, it is responsible for a drift of the particle inside a cylindrical cavity on time scales of the hour.

An idea could be constrain such a tiny force by considering a particle in an unstable inertial motion. An easy realization is to consider a charged particle inside a capacitor with its  two parallel walls normal to ${\bm e}_y$ with positions $y=\pm D$ and assume that there is a static electromagnetic field
$$
{\bm E}= E {\bm e}_y,\quad
{\bm B}= B {\bm e}_x.
$$
A particle launched with the velocity ${\bm V}_0= U{\bm e}_x$ will have a straight trajectory if 
\be
U=E/B.
\ee 
This is the standard classical Hall effect.

Now, assume there is a fifth force. The profile of the scalar field will be of the form $\phi(y)$ with $\partial_y\phi_0=0$ by symmetry. Hence, it implies, working with the non-relativistic equations of motion for the sake of simplicity, as
\begin{eqnarray}
 \ddot X &=& \frac{qB}{m A(\phi)} \dot Y,\\
 \ddot Y &=&\frac{qB}{m A(\phi)}\left(U-\dot X\right) - \frac{\beta}{M_{\rm P}}\partial_y\phi.
\end{eqnarray}
We rely of the computations of the profile of the scalar field we presented in Ref.~\cite{PhysRevD.100.084006}. Since $\phi\ll M_{\rm P}$ $A$ will almost not vary within the walls so that $A=A[(\phi(y=0)]\equiv 1$.

Then, consider a set of trajectories $\lbrace X(t;h),Y(t;h)\rbrace$ labeled by a parameter $h$, with initial conditions
$$
(X,Y)_0=(0,h),\qquad
(\dot X,\dot Y)_0 = (U,0).
$$
The trajectory $h=0$ will indeed be an inertial motion along $Y=0$ but, contrary to the usual Hall effect, the trajectories starting from $h\not=0$ will deviate from this standard trajectory.

Let us start by a toy profile mimicking the profile inside two walls, which has no analytic form,
\be
\phi(y) = \phi_{\rm wall} +\phi_0\left(1-\frac{y^2}{D^2}\right)
\ee
so that the force is
\be
{\bm F} = 2\frac{\phi_0\beta c^2}{M_{\rm P}D^2} y {\bm e}_y \equiv D\omega_0^2 \eta \frac{y}{D} {\bm e}_y
\ee
with $\eta=2(\phi_0/M_{\rm P})\beta c^2/D\omega_0^2\ll1$ the relative extra-acceleration induced by the fifth force.

 If the gradient is constant within the plates, which indeed not the case but allows to illustrate the phenomena, the trajectories are simply given by
\be
\left\lbrace
\begin{array}{ccc}
 X(t;h) &=& \left[ U + \frac{\eta}{1-\eta} h\omega_0 \right]t - \frac{\eta}{1-\eta} h \frac{\sin\sqrt{1-\eta} \omega_0 t}{\sqrt{1-\eta}}\\
 Y(t;h) &=& h\left[1+ \frac{\eta}{1-\eta} \left(1-\cos \sqrt{1-\eta} \omega_0 t \right)\right]
\end{array}
\right.
\ee
for $\eta<1$. We have the free parameters $U$ (determined by $E$ and $B$), $\omega_0$ (determined by $B$, the charge and mass of the particle), $h=1\ldots D$, $D$ determined by the size of the experiment so that then $\eta=F_0/D\omega_0^2$ is the quantity we want to constraint. Since we expect $F_0<10^{-7}$, $\eta$ is expected to be small compared to unity.

The main problem is that one would need extremely long capacitor which makes such an experiment completely unrealistic. One solution may be to consider periodic orbits and then turn to 2-dimensional configurations.

\end{document}